\newcommand{\Msolar}{\mbox{\,$\rm M_{\odot}$}}

\documentclass[usenatbib]{mn2e}

\usepackage{graphicx}
\usepackage{lscape}
\usepackage{txfonts}
\usepackage{url}

\begin{document}

\title[EW(H$\alpha$) as a proxy for sSFR]{The evolution
  of the equivalent width of the H$\alpha$ emission
  line and specific star-formation rate in star-forming galaxies at $\mathbf {1<z<5}$}
\author[M\'armol-Queralt\'o et al.]
{E. M\'armol-Queralt\'o$^{1}$\thanks{E-mail: emq@roe.ac.uk},
  R. J. McLure$^{1}$, F. Cullen$^{1}$, J. S. Dunlop$^{1}$, A. Fontana$^{2}$ and
\newauthor D. J. McLeod$^{1}$
\footnotesize\\
$^{1}$SUPA\thanks{Scottish Universities Physics Alliance}, Institute
for Astronomy, University of Edinburgh, Royal Observatory, Edinburgh EH9 3HJ\\
$^{2}$INAF$-$Osservatorio Astronomico di Roma, Via Frascati 33, Monte Porzio Catone, 00040, Rome, Italy}
\date{}
\pagerange{\pageref{firstpage}--\pageref{lastpage}} \pubyear{2015}
\def\LaTeX{L\kern-.36em\raise.3ex\hbox{a}\kern-.15em
    T\kern-.1667em\lower.7ex\hbox{E}\kern-.125emX}
\maketitle

\begin{abstract}
We present the results of a study which uses spectral energy
distribution (SED) fitting to investigate the evolution of the equivalent width (EW) of the H$\alpha$ emission line
in star-forming galaxies over the redshift interval $1<z<5$. 
After first demonstrating the ability of our SED-fitting technique 
to recover EW(H$\alpha$) using a sample of galaxies at $z\simeq 1.3$ with 
EW(H$\alpha$) measurements from 3D-HST grism spectroscopy, 
we proceed to apply our technique to samples of spectroscopically
confirmed and photometric-redshift selected star-forming galaxies at $z\geq 1$ 
in the CANDELS UDS and GOODS-S fields. Confining our analysis to a constant stellar 
mass range ($9.5<\log(M_{\star}/{\rm M}_{\odot})<10.5$), 
we find that the median EW(H$\alpha$) evolves only modestly with
redshift, reaching a rest-frame value of EW(H$\alpha$) $=301\pm{30}$\AA\, by 
redshift $z\simeq 4.5$. Furthermore, using estimates of star-formation rate (SFR) 
based on both UV luminosity and H$\alpha$ line flux, we use our galaxy samples
to compare the evolution of EW(H$\alpha$) and specific star-formation rate (sSFR). 
Our results indicate that over the redshift range $1<z<5$, the evolution displayed 
by EW(H$\alpha$) and sSFR is consistent, and can be adequately
parameterized as: $\propto (1+z)^{1.0\pm0.2}$. As a consequence, over this redshift 
range we find that the sSFR and rest-frame EW(H$\alpha$) of star-forming galaxies with stellar 
masses M$_{\star}$ $\simeq 10^{10}\Msolar$ are related by: 
EW(H$\alpha$)/\AA ~=~$(63\pm7)\times$~sSFR/Gyr$^{-1}$. Given the current 
uncertainties in measuring the SFRs of high-redshift galaxies, we conclude that 
EW(H$\alpha$) provides a useful independent tracer of sSFR for star-forming 
galaxies out to redshifts of $z=5$.
\end{abstract}

\section{Introduction}
Obtaining a full understanding of the physical processes underlying
the cosmic evolution of star formation and stellar mass assembly
remains a fundamental goal of extra-galactic astronomy.
Following the discovery of the so-called {\it main sequence} of star
formation \citep{Noeske2007,Daddi2007,Elbaz2007},
large amounts of observational effort have been invested in exploring the
form and evolution of the SFR$-M_{\star}$ relation 
\citep[e.g.][]{Karim2011,Whitaker2012,Whitaker2014, Speagle2014}.
Although notable disagreements concerning the normalisation and slope
of the main sequence still persist in the literature, it now seems
likely that these are dominated by selection biases and that, if dealt
with properly, the main sequence of star-forming galaxies has the form 
SFR $\propto M_{\star}^{0.7-1.0}$ out 
to redshifts of at least $z\simeq 3$ \citep[e.g.][]{Renzini-Peng2015,Johnston2015}, 
with a normalisation which mirrors the cosmic evolution of star-formation
rate density \citep[e.g.][]{Madau-Dickinson2014}.

At high redshifts ($z\geq 2$), much of the attention in the recent literature
has been focused on measuring the evolution of the specific star-formation rate 
(sSFR), defined as the ratio of SFR to stellar
mass \citep[e.g.][]{Gonzalez2014,deBarros2014,Tasca2015,Koprowski2015}. 
Although there is clear consensus that the sSFR of typical star-forming
galaxies rises rapidly from low redshift, reaching a value of $\simeq
2.5$ Gyr$^{-1}$ by $z\simeq 2$  (see Speagle et al. 2014 for a recent
review), the evolution of sSFR at higher redshifts has been much more controversial.
Initial studies \citep[e.g.][]{Stark2009,Gonzalez2010} indicated that for galaxies 
with stellar masses of $\simeq10^{10}\Msolar$, sSFR remains approximately constant 
at $\simeq 2.5$ Gyr$^{-1}$ over the redshift interval $2<z<6$. This result generated 
considerable interest, primarily because the apparent sSFR plateau is
difficult to reconcile with theoretical expectations that sSFR should
evolve $\propto (1+z)^{2.25}$, as it tracks the evolution of the
gas accretion rate onto dark matter halos \citep[e.g.][]{Neistein2008,Dave2011}.

Given the potentially large systematic uncertainties that can affect
stellar masses, and particularly SFRs, derived via SED fitting, 
it is obviously important to carefully consider the robustness of sSFR
measurements at high redshift. Indeed, one systematic uncertainty which
was not initially considered was the potential impact of nebular line
emission. At high redshifts ($z\geq 4$), measurements of key
stellar population parameters (i.e. age, stellar mass and star-formation
rate) are highly dependent on the strength of any photometric break
between the available near-IR and mid-IR photometry. 
Unfortunately, there is a basic degeneracy in how best
to model these photometric breaks with stellar population models. 
In most cases it is possible to obtain acceptable fits using mature (e.g. $\simeq
300$ Myr) stellar populations with moderate levels of star formation,
where a Balmer break is fitted between the near-IR and mid-IR photometry.
However, as indicated by \citet{Schaerer2009}, it is often possible
to obtain statistically identical (or improved) fits by invoking much
younger ($\leq 50$ Myr), lower mass stellar populations, with high
star-formation rates and strong nebular line emission. 
In these SED fits, breaks between the near-IR and mid-IR photometry
are typically the result of contamination of the mid-IR filters from strong 
[O{\sc iii}] or H$\alpha$ line emission. In some cases, this basic degeneracy can
lead to uncertainties in the estimated sSFR which are greater than an order of 
magnitude \citep[e.g.][]{Curtis-Lake2013}.

In a recent study, \citet{deBarros2014} performed SED
fitting on a sample of $\simeq$1700 Lyman-break galaxies (LBGs) at redshifts
$3<z<6$, using stellar population models which included nebular emission.
They concluded that $\simeq 65\%$ of LBGs display signs of significant
nebular line emission and that SED fits incorporating nebular emission
result in systematically lower stellar-mass estimates, and systematically
higher star-formation rate estimates. Although highly uncertain, 
\citet{deBarros2014} conclude that the sSFR evolves by a
factor of $5-50$ within the redshift interval $2<z<6$, more consistent
with the theoretical expectations (factor $\simeq 10$ increase) that the
results of \citet{Gonzalez2010}. However, when \citet{Gonzalez2014}
revisited the issue of how sSFR evolves at $z\geq 2$, they
arrived at a markedly different conclusion. Even after incorporating
nebular emission, \citet{Gonzalez2014} find that the sSFR of
galaxies with masses of $\sim5\times10^{9}\Msolar$ only increases by a factor
of $\simeq 2.3$ between $z=2$ and $z=6$, still in clear conflict with
theoretical expectations.

The key to resolving this issue is obtaining a reliable measurement of the
strength of nebular line emission in high-redshift galaxies. Within this
context, the evolution of EW(H$\alpha$) is of particular interest because, 
in principle, it should provide an independent method for determining the sSFR. 
This follows from the fact that EW(H$\alpha$) is the ratio of a star-formation
indicator (H$\alpha$ line flux) and a reasonable proxy for stellar mass
(stellar continuum at $\lambda_{rest}\simeq 6563$\AA). As a consequence,
it is credible to expect some level of consistency between the
redshift evolution of EW(H$\alpha$) and sSFR, although variation in the 
$M/L$ ratios of star-forming galaxies at the youngest ages and highest SFRs
could have an influence.

In fact, based on a combination of near-IR spectroscopy and narrow-band 
imaging, the recent literature presents a reasonably consistent picture of
how EW(H$\alpha$) evolves out to $z\simeq 2$ 
\citep[e.g.][]{Erb2006,Fumagalli2012,Kashino2013,Sobral2014}.
However, the situation at higher redshifts is much less clear. 
Due to the lack of available mid-IR spectroscopy, at $z\geq 2.5$ it is
currently only possible to measure EW(H$\alpha$) via the excess
emission detected in the {\it Spitzer} IRAC 3.6$\mu$m and 4.5$\mu$m filters.
In particular, the redshift interval $3.8<z<5.0$ represents a
sweet-spot, where H$\alpha$ contaminates the 3.6$\mu$m filter but the 4.5$\mu$m 
filter remains line free and suitable for anchoring SED-based continuum estimates. 

Two previous studies have adopted an SED-based approach to measuring EW(H$\alpha$) 
at $3.8<z<5.0$. Firstly, \citet{Shim2011} derived EW(H$\alpha$) measurements based 
on SED fits to the optical-mid-IR photometry for a final sample of 64 
spectroscopically-confirmed star-forming galaxies at $3.8<z<5.0$, {\it excluding} 
the contaminated 3.6$\mu$m photometry from their SED fits. 
The results of the \citet{Shim2011} study suggested that galaxies with
stellar masses of $\simeq 4\times10^{9}\Msolar$ (converted to Chabrier IMF),
have a median EW(H$\alpha$) $\geq 600$\AA, a factor of $\simeq 3$ larger than 
EW(H$\alpha$) measurements at $z\simeq 2$ derived from near-IR spectroscopy 
(e.g. Erb et al. 2006). More recently, \citet{Stark2013} revisited this issue, 
using a final sample of 45 spectroscopically confirmed galaxies in the redshift
interval $3.8<z<5.0$ (30 in common with Shim et al. 2011). Depending on whether 
they included or excluded the contaminated 3.6$\mu$m photometry in their SED 
fitting, \citet{Stark2013} found that the typical value of EW(H$\alpha$) in 
their sample was in the range $270-410$\AA. 

Motivated by the continuing uncertainty over the evolution of
EW(H$\alpha$), and how it relates to the evolution of the
sSFR, in this paper we use SED fitting to consistently explore the evolution
of EW(H$\alpha$) and sSFR over the redshift interval $1<z<5$. 
To achieve this, we analyse samples of spectroscopically-confirmed
star-forming galaxies at $1.20<z<1.50$, $2.10<z<2.45$ and
$3.8<z<5.0$ in the CANDELS regions of the UDS and GOODS-S fields.
In addition, we re-enforce our measurements of EW(H$\alpha$) at $z\geq 4$ by 
analysing a much larger sample of photometric-redshift selected
star-forming galaxies at $3.8<z_{phot}<5.0$. Within these three redshift
windows the H$\alpha$ emission line contaminates the {\it HST} F160W (hereafter
$H_{160}$), $K_{s}$ and 3.6$\mu$m filters respectively. Consequently, it is clear 
that to accurately measure EW(H$\alpha$) at
$z\geq 1$ based on SED-fitting requires deep $K-$band data in
combination with deep {\it and} accurately deconfused 3.6$\mu$m+4.5$\mu$m 
photometry.

Within this context, our study makes use of important new data-sets
that were previously unavailable. Firstly, we exploit the photometry from the
aperture-matched catalogues of the CANDELS UDS and GOODS-S fields
\citep{Galametz2013,Guo2013}, which feature accurately deconfused IRAC fluxes 
extracted using the available $H_{160}-$band imaging as a high-resolution prior. 
Secondly, the UDS and GOODS-S CANDELS fields have recently been imaged to 
$5\sigma$ depths of $25.5-26.2$(AB) in the $K_{s}-$band by the HUGS survey 
\citep{HUGS}. Thirdly, we use our own reductions of the 3D-HST \citep{Brammer2012} 
grism survey data covering the UDS and GOODS-S fields. In addition to providing 
spectroscopically-confirmed star-forming galaxies within the traditional redshift 
desert ($1.5<z<2.5$), the 3D-HST data provides a sample of star-forming galaxies at 
$1.2<z<1.5$ with direct spectroscopic measurements of EW(H$\alpha$) with which to 
validate our SED-fitting technique.

The structure of the paper is as follows. In Section 2 we describe the photometric 
and spectroscopic data-sets used in our analysis and our criteria for selecting a 
uniform sample of star-forming galaxies. In Section 3 we describe our SED-fitting 
technique for determining EW(H$\alpha$) and our adopted prescriptions for 
calculating stellar masses and SFRs. In Section~\ref{sec:testing_method} 
we describe the validation and
calibration of our SED-fitting technique based on the 3D-HST grism spectroscopy. 
In Section~\ref{sec:results} we present our new EW(H$\alpha$) and sSFR results and 
compare them to 
recent studies in the literature. In Section 6 we explore the combined evolution of 
EW(H$\alpha$) and sSFR and discuss whether or not EW(H$\alpha$) can be used as a 
useful proxy for sSFR at high redshift. In Section 7 we present a summary of our 
main results and conclusions. Throughout the paper we use the AB magnitude system 
\citep{AB_system1,AB_system2} and adopt the following cosmology: $\Omega_{\rm m}
= 0.3$, $\Omega_{\rm \Lambda} = 0.7$ and H$_0 = 70$ km~s$^{-1}$~Mpc$^{-1}$. Unless 
otherwise stated, we use EW(H$\alpha$) to refer to the rest-frame equivalent width 
throughout.

\begin{table*}\label{table_properties}
\caption{Summary of the measured and derived properties of the four different galaxy 
samples analysed in this study. Column one lists the sample name and column two 
lists the number of objects within each sample. The final sample in the table is 
simply the combination of the spectroscopic and photometric-redshift selected 
samples at $z\simeq 4.5$. Columns three and four list the median redshift and 
median stellar mass for each sample. Column five lists the median rest-frame EW of 
the H$\alpha$ emission line as derived from our SED fitting. Columns six and seven 
list the median values of sSFR, as derived from the rest-frame UV and the measured 
EW(H$\alpha$) respectively (see text for full details).}
\begin{tabular}{lcccccc}
\hline
\hline
Sample & N & z & $\log(M_{\star}/\Msolar$) & EW(H$\alpha$)/\AA & sSFR$\_$UV /Gyr$^{-1}$ & sSFR$\_$H$\alpha$/Gyr$^{-1}$ \\
\hline
spec-z ($1.2<z<1.5$) &            143 & 1.34& 9.83 & $146\pm\phantom{0}9$ & $2.0\pm0.2$ & $1.7\pm0.2$\\

spec-z ($2.1<z<2.5$) &  \phantom{0}71 & 2.25& 9.77 & $217\pm17$     & $4.0\pm0.3$& $4.5\pm0.3$\\

spec-z ($3.8<z<5.0$) &  \phantom{0}26 & 4.55& 9.79 & $288\pm92$     &$4.3\pm0.6$ & $5.5\pm1.1$\\

phot-z ($3.8<z<5.0$) &           129  & 4.34& 9.69 & $313\pm29$     &$5.4\pm0.3$ & $5.1\pm0.6$\\

full\hspace{9pt} ($3.8<z<5.0$) &           155  & 4.38& 9.70 & $301\pm30$     &$5.3\pm0.3$ & $5.4\pm0.5$\\

\hline
\end{tabular}
\end{table*}

\section{Data}\label{sec:data}
The analysis in this paper is based primarily on the photometry and 
spectroscopy of the UDS and GOODS-S fields provided by the public Cosmic Assembly 
Near-IR Deep Extragalactic Legacy Survey 
\citep[CANDELS,][]{Grogin2011, Koekemoer2011} and 3D-HST spectroscopic
surveys \citep{Brammer2012} respectively. In common with all five of the CANDELS 
survey fields, the UDS and GOODS-S feature the deep optical, near-IR and 
{\it Spitzer} IRAC imaging necessary to trace the evolution of the EW(H$\alpha$) 
via SED fitting. However, crucially, amongst the CANDELS fields it is {\it only} 
the UDS and GOODS-S which feature suitably deep $K_{s}-$band imaging, provided by
the recently completed HUGS survey \citep{HUGS}.

\begin{figure*}
\centering
\includegraphics[width=0.8\textwidth,clip=true]{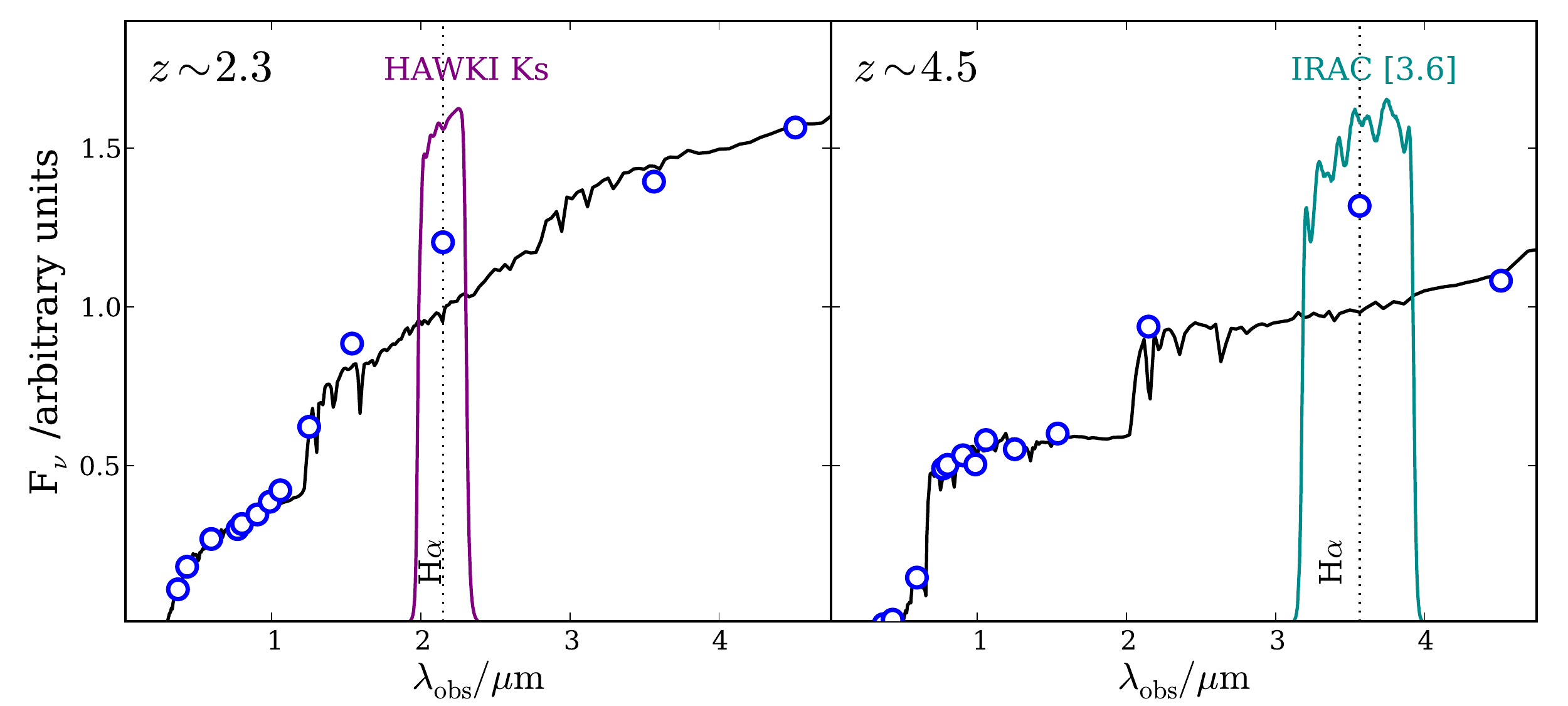}
\caption{The left-hand panel shows the stacked multi-wavelength
  photometry for the galaxies in the $2.1<z<2.45$ spectroscopic
  sample drawn from the GOODS-S field. The solid black line shows a
  stack of the best-fitting SED templates (derived by excluding
  filters contaminated by strong emission lines from the fitting process).
  The purple line indicates the transmission profile of the $K_{s}-$band filter, 
  which should be contaminated by the H$\alpha$ emission line at these
  redshifts. The stacked $K_{s}-$band photometry shows a clear excess in comparison
  to the stacked SED templates (it can also be seen that the $H_{160}-$ filter is contaminated by
  [O{\sc iii}] emission). The right-hand panel shows the
  equivalent information for the $z\simeq 4.5$ spectroscopic galaxy
  sample. In this case, a clear flux excess due to H$\alpha$ emission can be seen in the 
  IRAC $3.6\mu$m filter (turquoise line).}
\label{fig:phot_stack}
\end{figure*}

\subsection{CANDELS photometry}
The success of this study relies on the quality of the available photometry, and for
our purposes we have adopted the publicly available photometry catalogues of the 
CANDELS UDS and GOODS-S fields published by \cite{Galametz2013} and \cite{Guo2013}.
For the GOODS-S field we make use of the update to the \cite{Guo2013}
catalogue released by \cite{HUGS} which includes the final $K_{s}-$band photometry from the HUGS survey.
Both catalogues are selected using the CANDELS $H_{160}$ imaging and 
feature PSF-matched isophotal photometry measured from the {\it  HST} ACS and 
WFC3/IR imaging available across both fields. Likewise, both catalogues make use of 
the {\sc tfit} \citep{Laidler2007} deconfusion package which utilises $H_{160}$ 
priors to obtain aperture-matched photometry in the available optical+nearIR 
ground-based imaging and {\it Spitzer} IRAC data. As highlighted above, the latest 
versions of both catalogues feature {\sc tfit}-generated photometry in the 
$K_{s}-$band from the recently completed HUGS survey. Full details of how the 
photometry catalogues were generated can be found in \cite{Galametz2013} and 
\cite{Guo2013} respectively.


\subsection{3D-HST spectroscopy}
The 3D-HST survey provides low spectral resolution, spatially resolved, near-IR 
grism spectroscopy of all five of the CANDELS survey fields \citep{Brammer2012}. 
Specifically, the 3D-HST survey employs the G141 grism on WFC3/IR to provide spectra with 
$R\sim130$ over the wavelength range $1.10 - 1.68 \mu$m. The raw data from 3D-HST 
are publicly available, and we have employed our own modified version of the 
{\sc axe} software package \citep{Kummel2009} to reduce the data available over the 
UDS and GOODS-S fields. The grism spectra we employ here were originally
reduced for the gas-phase metallicity study of \citet{Cullen2014},
to which the reader is referred for further details of the reduction
and redshift determination processes.

The 3D-HST spectra play a crucial role in validating our SED-fitting technique for 
measuring EW(H$\alpha$). In the redshift interval $1.2<z<1.5$, where the H$\alpha$
emission-line contaminates the $H_{160}-$filter, the 3D-HST spectra cover both the 
[O{\sc iii}] and H$\alpha$ emission lines. Consequently, within this redshift range, 
the 3D-HST spectra can provide both unambiguous spectroscopic redshifts and 
{\it direct} spectroscopic measurements of EW(H$\alpha$). In addition, the 3D-HST 
spectra are an excellent resource for obtaining emission-line redshifts of 
star-forming galaxies in the traditional redshift desert and provide many of the 
spectroscopically-confirmed objects in our $z\simeq 2.3$ sub-sample where H$\alpha$ 
contaminates the $K_{S}-$filter. 

\subsection{Spectroscopic galaxy sample}
In order to study the evolution of EW(H$\alpha$) at $z>1$, we have assembled 
samples of spectroscopically-confirmed star-forming galaxies in the redshift 
ranges: $1.2<z<1.5$, $2.1<z<2.45$ and $3.8<z<5.0$, where the H$\alpha$ emission 
line contaminates the $H_{160}$, $K_{s}$ and IRAC 3.6$\mu$m filters respectively. 
In addition to galaxies drawn from the 3D-HST survey, the initial samples were 
drawn from the various different spectroscopic studies of the GOODS-S and UDS 
fields: \citet{Cimatti2002}, \citet{VVDS}, \citet{Mignoli2005}, 
\citet{Vanzella2008}, \citet{Popesso2009}, \citet{Cooper2012}, \citet{McLure2013} 
and \citet{Morris2015}. Objects were only considered for selection if they had 
the highest-quality spectroscopic redshift flags.

After the initial selection process, the final samples were restricted to those 
star-forming galaxies for which statistically acceptable SED fits were 
obtained at their spectroscopic redshifts, with corresponding stellar mass 
measurements lying in the range $9.5<\log(M_{\star}/\Msolar)<10.5$ (see 
Section 3). Moreover, in order to exclude passive/quiescent systems, galaxies 
with
UV-luminosity based SFR estimates inconsistent with lying on the main sequence were also excluded.
Although our final spectroscopic galaxy samples are clearly not complete, their
location on the main sequence (see Section 5.1) demonstrates that they are fully
representative of the dominant star-forming galaxy population at $1<z<5$  
in the mass range $9.5<\log(M_{\star}/\Msolar)<10.5$. The basic properties of 
the three spectroscopic star-forming galaxy samples are provided in 
Table~\ref{table_properties}.

\subsection{Photometric galaxy sample}
The two lower-redshift galaxy samples at $z\simeq 1.3$ and $z\simeq 2.3$ are 
selected within sufficiently narrow redshift intervals (i.e. $\Delta z=0.3$ and 
$\Delta z=0.35$ respectively) that the necessity for spectroscopic redshifts is 
clear. However, the width of the IRAC 3.6$\mu$m filter allows us to study the impact
of H$\alpha$ emission over the redshift range $3.8<z<5.0$ (i.e., $\Delta z=1.2$).
Moreover, the presence of a strong Lyman-break in their spectra makes the selection
of star-forming galaxies at these redshifts reasonably straightforward.

Consequently, in order to boost the statistics in our highest-redshift bin, we 
assembled a sample of Lyman-break galaxies based on our own photometric redshift 
analysis of the \citet{HUGS} and \citet{Galametz2013} catalogues. Applying 
the same stellar mass and sSFR criteria (and excluding the objects in common with 
the $z\simeq 4-5$ spectroscopic sample) the final sample of photometric-redshift 
selected galaxies comprises 129 objects in the redshift range $3.8<z_{photo}<5.0$. 
The basic properties of the final photometric sample of star-forming galaxies are 
also shown in Table~\ref{table_properties}. The good
agreement between the median properties inferred for the photometric and 
spectroscopic samples of galaxies at $3.8<z<4.5$ is notable, confirming that the spectroscopic
sample is not a biased sub-set of star-forming galaxies in this redshift 
range. For that reason, from now on we will quote the values for the full sample (i.e., 
phot-z + spec-z) when referring to this redshift range.


\section{Equivalent widths, stellar masses and star-formation rates}
In this section we describe our SED-fitting technique for measuring 
EW(H$\alpha$) from the available multi-wavelength photometry.  
We also describe our adopted prescriptions for calculating the stellar masses and 
star-formation rates of the galaxies in our spectroscopic and photometric samples. 

\subsection{SED fitting}
The photometry for all objects included in this study was analysed
using the template-fitting software described in \citet{McLure2011}. 
A fuller description of the latest version of this software is 
provided in \citet{McLeod2015}, but we provide the essential details
here for completeness. The standard version of the code uses \citet{BC03}
(hereafter BC03) SED templates, combined with the
\citet{Calzetti2000} dust attenuation law and the \citet{Madau1995} prescription 
for IGM absorption. Strong nebular emission lines can be included in the template 
fitting, with the H$\alpha$ line flux calculated from the template star-formation
rate and the strength of other significant nebular lines set using the
line ratios determined by \citet{Cullen2014}. The template fitting is performed in 
flux-space to the allow the proper treatment of flux errors and, in addition to 
photometric redshifts, the code delivers best-fitting values of stellar mass, 
star-formation rate and synthetic photometry for the best-fitting template.

Throughout the SED-fitting process we adopted BC03 stellar evolutionary models with 
solar and sub-solar metallicities ($Z_{\odot}, 0.5Z_{\odot}$ \& $0.2Z_{\odot}$), a 
Chabrier initial mass function (IMF) and a range of star-formation histories with 
exponentially decaying SFRs ($\tau$-models) in the range $0.2 <\tau < 10$ Gyr.
In obtaining the best-fitting SED, dust attenuation was allowed to vary over the 
range $0<A_{V}<2.5$ and template ages were required to be between 50 Myr and the 
age of the Universe at a given redshift.

\begin{figure*}
\centering
\includegraphics[width=1.0\textwidth,clip=true]{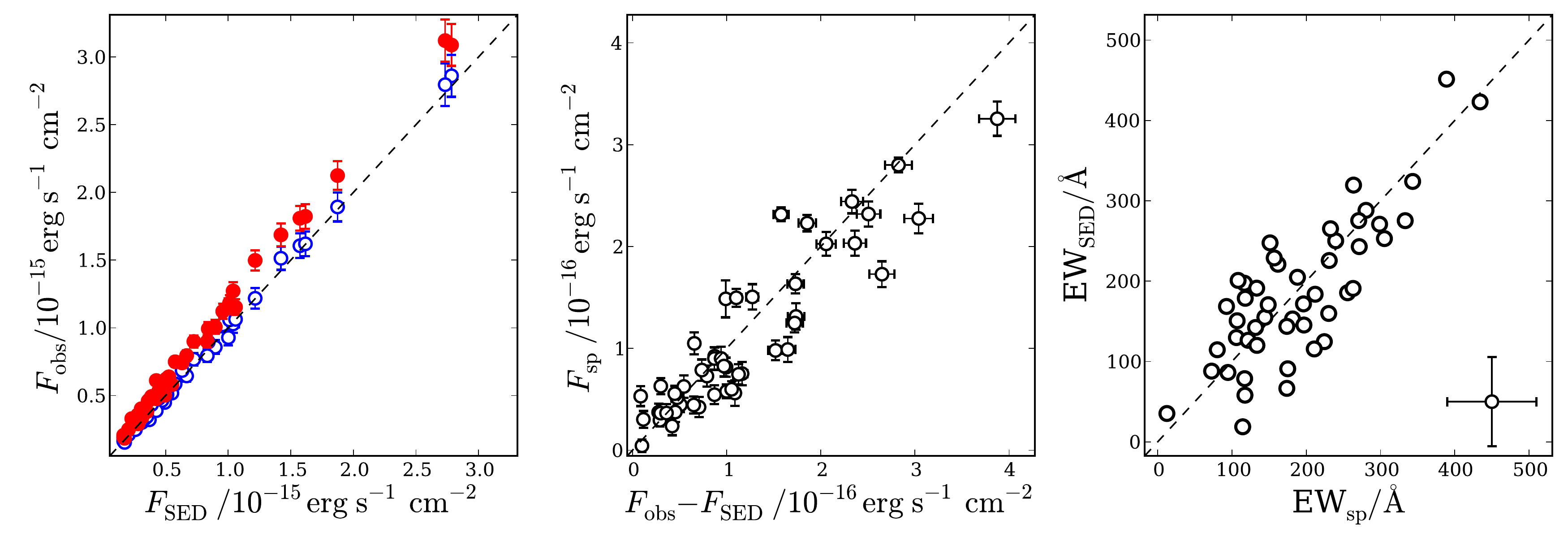}
\caption{Each panel shows data for the final sample of 48 galaxies at $z\simeq 1.3$ 
for which it was possible to extract reliable measurements of EW(H$\alpha$) from 
their 3D-HST spectra. In each panel the dashed line indicates a 1:1 relation. The 
left-hand panel compares the average continuum flux calculated from the observed 
$H_{160}$ magnitude (filled red circles) with the continuum flux predicted by SED 
fitting of the multi-wavelength photometry, excluding filters contaminated by strong 
emission lines. The open blue data points show the continuum flux measured from the 
observed $H_{160}$ magnitude, after subtracting the H$\alpha$ line flux measured 
from the 3D-HST spectra. The middle panel shows the H$\alpha$ line flux as measured 
from the spectra versus the line flux predicted from the difference between the 
observed and synthetic $H_{160}$ magnitudes. As expected, the predicted line flux
is $\simeq 20\%$ larger than the measured H$\alpha$ line flux, simply because the 
raw output from the SED fitting is a measurement of the flux of all emission lines 
within the H$_{160}$ filter (i.e. H$\alpha$+[N{\sc ii}]+[S{\sc ii}]).
The right-hand panel shows EW(H$\alpha$) derived from the SED fitting
versus EW(H$\alpha$) measured from the 3D-HST spectra. 
Under the assumption that there should exist a 1:1 relation between the two, the EW$_{\rm{SED}}$ values have been 
scaled by a factor of $f=0.9$ (see text for details). A representative error bar, 
computed as the median value of the individual error bars for each galaxy, is 
shown in the lower right-hand corner.}
\label{fig:testing_z1}
\end{figure*}

\subsection{Equivalent width and stellar mass}
Two separate SED fits were performed for each object, in both cases fixing the 
redshift to the spectroscopic value (or best-fitting $z_{phot}$ value for
the photometrically selected $z\simeq 4.5$ sample) and adopting the same set of SED templates.

In the first SED fit all of the available multi-wavelength photometry
was included and nebular emission lines were added to the SED templates.
It is the best-fitting values of stellar mass returned by these SED
fits which are adopted throughout the subsequent analysis.
In contrast, during the second SED fit all filters potentially
contaminated by a strong nebular emission lines (i.e. [O{\sc ii}],
[O{\sc iii}] or H$\alpha$) were {\it excluded} from the fit. The results
of this SED fit are used to calculate the rest-frame EW of the
H$\alpha$ emission line via the formula:
\begin{equation}
{EW  = \frac{W_{rec}}{(1+z)}\big( 10^{(-0.4\Delta mag)} -1 \big)}
\end{equation}
\noindent
where $W_{rec}$ is the rectangular width of the filter contaminated by
H$\alpha$ and $\Delta mag$ is the difference between the observed
magnitude in that filter and the synthetic magnitude from the SED-fit
excluding filters potentially contaminated by strong line emission
(i.e. $\Delta mag=m_{obs}-m_{SED}$). 
As previously discussed, in this 
study we consider H$\alpha$ contamination of the $H_{160}$, $K_{s}$ 
and IRAC 3.6$\mu$m filters for galaxies at $z\simeq 1.3$, $z\simeq 2.3$ and 
$z\simeq 4.5$. For these three filters we adopt $W_{rec}$ values of 2683\AA,
3150\AA\, and 6844\AA\, respectively. In reality, the SED fits return a measurement of the total EW of
all the emission lines within the relevant filter (i.e. H$\alpha$,
[N{\sc ii}] and [S{\sc ii}]). In Section~\ref{sec:testing_method} we 
calibrate for this effect by comparing our SED-based results with direct 
spectroscopic EW measurements.

As an illustration of how the SED-fitting method works, Fig.~\ref{fig:phot_stack} shows the stacked 
multi-wavelength photometry and the stacked best-fitting SED 
templates for the spectroscopically-confirmed galaxies at $z\simeq2.3$ (left panel) 
and $z\simeq4.5$ (right panel) in the GOODS-S field. Compared to the
best-fitting SED templates (fit by excluding filters potentially
contaminated by line emission), the excess flux in the $K_{s}$ and
IRAC 3.6$\mu$m filters, caused by H$\alpha$ emission at $z\simeq 2.3$
and $z\simeq 4.5$ respectively, is clearly visible.


\subsection{Star formation rate}
When combined with spectroscopic redshifts, the high $S/N$ ratio UV-MidIR data 
available within the UDS and GOODS-S CANDELS fields provides SED-based stellar 
mass measurements which are relatively well constrained 
\citep[e.g.][]{Mobasher2015}. Indeed, the median stellar masses of our four galaxy 
samples (see Table~\ref{table_properties}) are stable at the $\pm 0.15$ dex level, 
irrespective of the allowed range in metallicity, reddening and star-formation 
timescale (or indeed, whether filters contaminated by line emission are included or 
excluded).

Unfortunately, however, the estimates of star-formation rate derived from SED
fitting are far less stable, showing large discrepancies depending on the adopted 
dust reddening, metallicity and star-formation histories. As a result, adopting 
different modelling assumptions, it is perfectly possible to obtain statistically 
acceptable SED fits to our samples of $1<z<5$ star-forming galaxies which result in 
median sSFR values which differ by a factor of $\simeq 4$. Consequently, throughout 
the analysis presented in this paper we have adopted two different, semi-empirical, 
estimates of star-formation rate, both of which have the benefit of being largely 
independent of the adopted set of SED templates. 

\subsubsection{UV star-formation rate estimate}
Our primary star-formation rate estimate is based on the far-UV luminosity
of each galaxy, as measured from the best-fitting SED template (excluding 
line-contaminated filters) using a 100\AA-wide top-hat filter centred on 
$\lambda_{rest}=1500$\AA. To calculate the reddening we have derived the UV spectral
slope ($\beta$) of each galaxy using top-hat filters centred on $1410$\AA\, and 
$2400\AA$, corresponding to UV windows no. 4 and 10 from \citet{Calzetti1994}. 
Based on the derived values of $\beta$, the dust reddening was then calculated 
using the \citet{Meurer1999} correlation between dust extinction (A$_{1600}$) and 
$\beta$. Once the 1500\AA\, luminosities were dust corrected (assuming
A$_{1500}=1.04$A$_{1600}$, as predicted by the \citet{Calzetti2000} attenuation 
law), the final star-formation rate estimates were derived using the updated 
correlation between UV luminosity and SFR adopted by \citet{Madau-Dickinson2014}:
\begin{equation}
\log({\rm SFR}) = \log(L_{1500}) - 28.14
\end{equation}
\noindent
where SFR is measured in $\Msolar$ yr$^{-1}$, $L_{1500}$ is measured in 
erg~s$^{-1}$~Hz$^{-1}$ and we have corrected from a Salpeter to a Chabrier IMF
assuming a correction factor of 0.63.

\subsubsection{H$\alpha$ star-formation rate estimate}
The second star-formation rate estimate is based on the H$\alpha$
line flux of each galaxy, as derived from the corresponding EW(H$\alpha$) 
measurement. In performing this calculation, the continuum flux is estimated from 
the best-fitting SED template (excluding line-contaminated filters) using a 
100\AA-wide top-hat filter centred on $\lambda_{rest}=6563$\AA. Moreover, the 
H$\alpha$ fluxes are dereddened assuming A$_{6563}$=0.33A$_{1600}$, as predicted by 
the \citet{Calzetti2000} reddening law where, as before,  A$_{1600}$ is derived 
using the A$_{1600}-\beta$ correlation from \citet{Meurer1999}. This calculation 
therefore explicitly assumes that the nebular and continuum reddening are identical, 
as found in a recent study of 3D-HST star-forming galaxies at $z\simeq 2.2$ by 
\citet{Cullen2014}, and consistent with the recent work by \citet{Reddy2015} for galaxies
with our median mass and SFR. We have adopted the following relationship between H$\alpha$ 
luminosity and SFR:
\begin{equation}
\log({\rm SFR}) = \log(L_{\rm{H}\alpha}) - 41.35 
\end{equation}
\noindent
where SFR is measured in $\Msolar$ yr$^{-1}$ and $L_{{\rm H}\alpha}$ is measured 
in erg s$^{-1}$. We note that this calibration produces SFR estimates $\sim 10\%$ 
lower than the recent calibration of \citet{Kennicutt-Evans2012} and $\sim 10\%$ 
higher than the calibration of \citet{Madau1998}, when both are corrected to a 
Chabrier IMF.

\begin{figure}
\centering
\includegraphics[width=0.48\textwidth,clip=true]{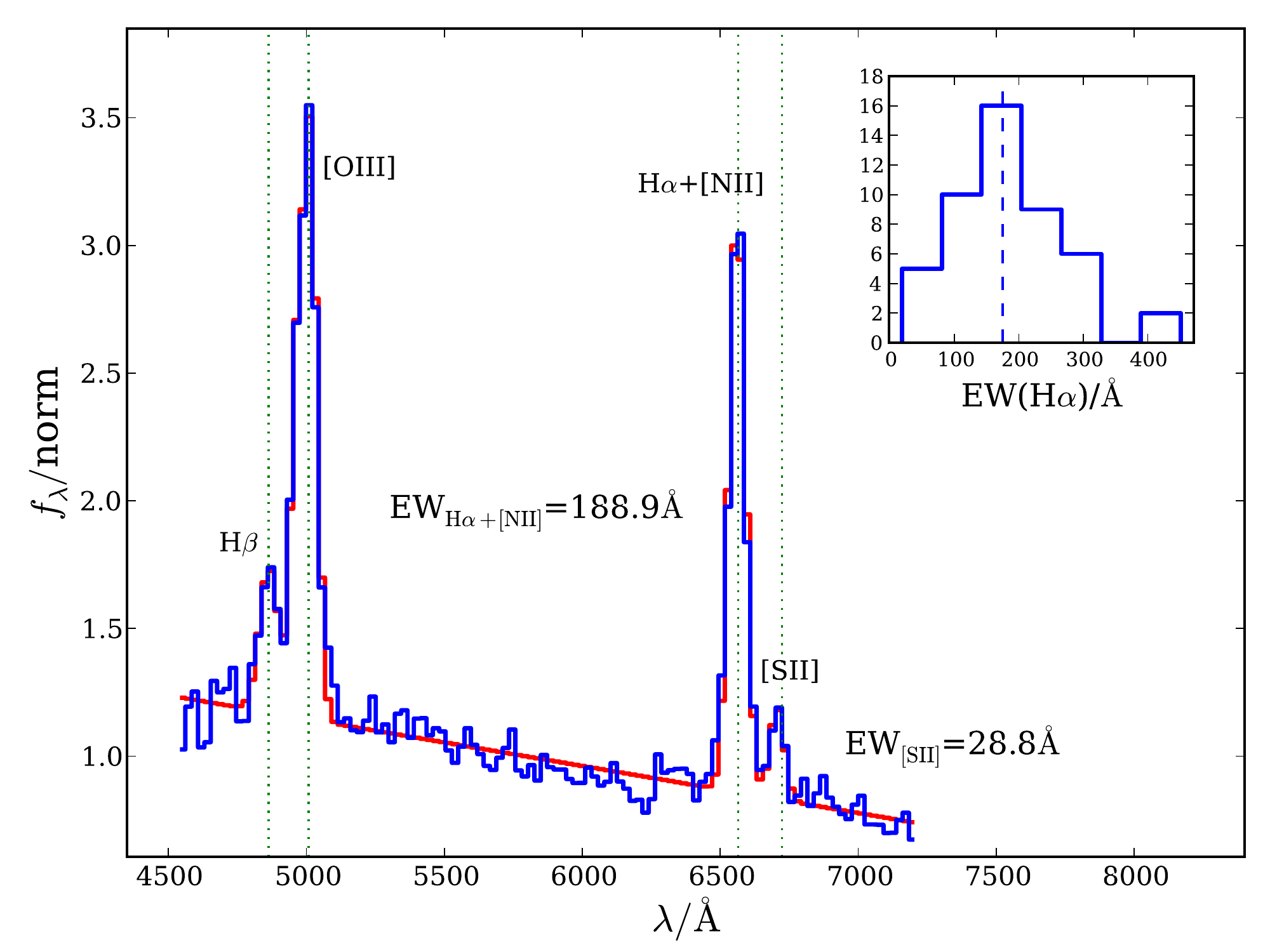}
\caption{The median stack of the 3D-HST spectra for the 48 galaxies at
  $z\simeq 1.3$ featured in Fig.~\ref{fig:testing_z1} is shown in blue. The red line
  shows the best-fitting continuum+emission line model.
The measured value of EW(H$\alpha$+[N{\sc ii}])  from the
stacked spectrum is $189\pm19$\AA. Assuming that H$\alpha$ contributes
90\% of the H$\alpha$+[N{\sc ii}] flux \citep{Sanders2015} the stacked spectrum
therefore indicates a median EW(H$\alpha$) of $170\pm17$\AA. 
The inset shows the distribution of EW(H$\alpha$) measurements from
our SED fitting of the individual objects that went into the
stack. The median value of the individual EW(H$\alpha$) measurements returned by the SED fitting is $175\pm14$\AA, in
excellent agreement with the value measured from the stacked spectra.}
\label{fig:sp_stack}
\end{figure}

\section{Testing the SED-fitting method}\label{sec:testing_method}
Before proceeding to present the main results, in this section we provide details 
of the tests which were performed to validate and calibrate our SED-fitting 
technique for measuring EW(H$\alpha$). All of the tests were performed using a 
sample of star-forming galaxies at $z\simeq 1.3$, drawn from the 3D-HST sample in 
the UDS and GOODS-S fields. As previously discussed, these galaxies are ideal for 
our purposes because they allow a comparison of the SED-based EW measurements with 
direct measurements from the 3D-HST spectra.

\begin{figure*}
\centering
\includegraphics[width=0.75\textwidth,clip=true]{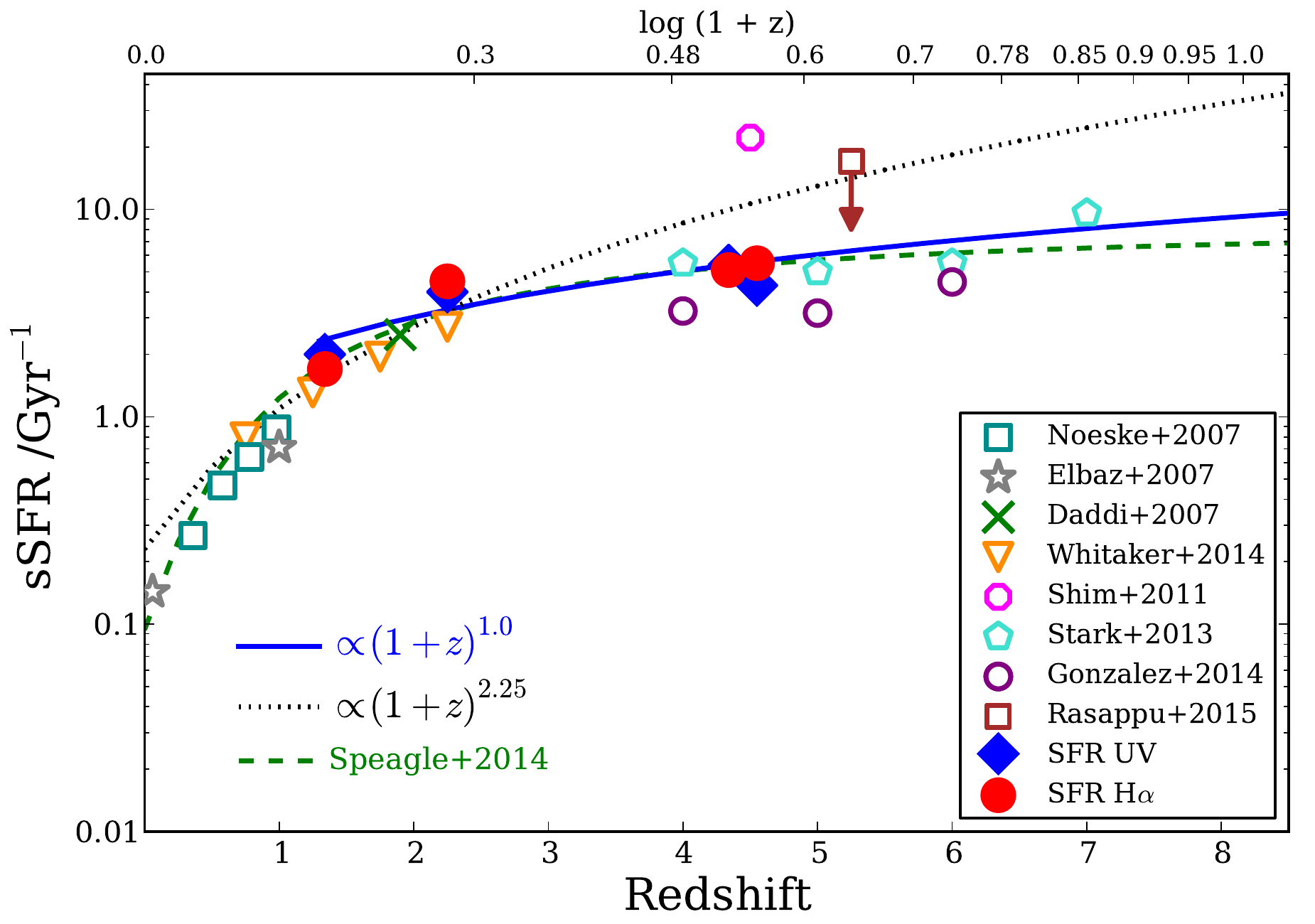}
\caption{The evolution of sSFR as a function of redshift. The filled
blue diamonds and red circles show the median values of sSFR for 
the four galaxy samples presented in Table~\ref{table_properties}. The filled 
blue diamonds show our median sSFR measurements based on UV luminosity, while 
the filled red circles show the median sSFR measurements as derived from our 
SED-based measurements of EW(H$\alpha$). 
The results of various different literature studies have also been plotted for
comparison. The dotted black line shows sSFR evolution of the
form: $\propto (1+z)^{2.25}$, as typically predicted by galaxy
evolution models (normalised to pass through the cluster of
data points at $1<z<2$). The dashed green line shows the sSFR$-z$ relation
derived by \citet{Speagle2014} for galaxies with stellar mass
$\log(M_{\star}/\Msolar)=9.8$, the same as the median mass of our galaxy samples. 
The blue solid line shows a relation of the form: $\propto (1+z)^{1.0}$,
which was fit to our new data points alone (see text for discussion).
It can be seen that our sSFR results, based on two independent measurements, are 
both internally consistent and in excellent agreement with the sSFR$-z$ relation 
derived by \citet{Speagle2014} for galaxies of this mass.}
\label{fig:evolution_sSFR}
\end{figure*}
\subsection{Individual 3D-HST spectra}
The H$\alpha$ emission line lies securely within the $H_{160}$-filter for galaxies 
within the redshift interval $1.2<z<1.5$. Consequently, our initial sample consisted 
of all galaxies within this redshift range, with high $S/N$ ratio detections of the 
H$\alpha$ emission line. In addition, we also required that the 3D-HST spectra 
provided unambiguous spectroscopic redshifts (effectively meaning that 
[O{\sc iii}]5007 was also detected) and a continuum detection free from significant 
contamination due to overlapping spectra from nearby objects. After applying these 
criteria the final test sample consisted of 48 galaxies with redshifts in the 
range $1.23<z<1.49$.

The individual spectra were fitted with a combination of a linear
continuum and two Gaussians to reproduce the blended [O{\sc iii}] doublet, 
H$\alpha$ and [N{\sc ii}] emission lines. After subtracting the continuum, the
fluxes of the blended H$\alpha$+[N{\sc ii}] emission lines were measured
within $\pm 3 \sigma$ from the centroid of the best-fitting Gaussian,
and EW(H$\alpha$) was calculated under the assumption that H$\alpha$ 
contributes 90\% of the total H$\alpha$+[N{\sc ii}] flux \citep[taken from][for 
star-forming galaxies with similar SFRs in our stellar mass range]{Sanders2015}.
To estimate the error in derived EW measurements a set of 100 realisations was run,
in which the 3D-HST spectrum is perturbed according to its error spectrum.

The results of the SED-fitting tests on the final sample of 48 test galaxies at 
$z\simeq 1.3$ are shown in Fig.~\ref{fig:testing_z1}. In the left-hand panel 
the average continuum 
flux calculated from the observed $H_{160}$ magnitude (red data points) is plotted 
against the continuum flux calculated from the synthetic $H_{160}$ magnitude 
returned by the best-fitting SED template (excluding filters potentially 
contaminated by line emission). As expected, a clear flux excess is apparent. In 
contrast, the blue data points show the average continuum flux after the observed 
$H_{160}$ photometry has been corrected by subtracting the emission-line flux
measured from the 3D-HST spectra. As can clearly be seen, the two continuum 
estimates are now in excellent agreement. The middle panel of 
Fig.~\ref{fig:testing_z1} shows the H$\alpha$ line flux measured from the 3D-HST
spectra versus the H$\alpha$ line flux estimated from the difference between the 
observed and synthetic $H_{160}$ photometry. It can be seen that the two independent
H$\alpha$ line flux measurements are well correlated, with a tendency for the
SED-based estimate to be systematically $\simeq 20\%$ higher. Again, this is as 
expected, given that the SED-based results inevitably yield an estimate of the 
total H$\alpha$+[N{\sc ii}]+[S{\sc ii}] line flux. 

Finally, the right-hand panel of Fig.~\ref{fig:testing_z1} shows EW(H$\alpha$) 
measured from the SED-fitting, versus EW(H$\alpha$) measured from the 3D-HST 
spectra. To produce the 
final calibration of the relationship between the SED-based EW measurements 
and the direct spectroscopic measurements, we have explicitly assumed that there 
should exist a 1:1 relation between the two, and that the only freedom should be to 
introduce a multiplicative factor ($f$) in order to correct the SED-based
measurements for the additional flux of the [N{\sc ii}] and [S{\sc  ii}] emission 
lines. Under this assumption, it was found that scaling the SED-based EW
measurements by $f=0.9$ produced the best reproduction of the EW measurements from 
the 3D-HST spectra. As demonstrated by the right-hand panel of 
Fig.~\ref{fig:testing_z1}, based on this
scaling, the SED-based and spectroscopically measured EWs follow a 1:1 relation. 
Unless otherwise stated, all EW(H$\alpha$) measurements based on
SED-fitting quoted in this paper have been scaled by $f=0.9$ according to this calibration.

\subsection{Stacked 3D-HST spectra}
Having demonstrated the validity of our SED-based technique for
measuring EW(H$\alpha$) on individual objects, an additional test
was performed on the median stack of the 48 spectra from the test sample.
For those 48 galaxies, we first normalized their spectrum to the region 
$\lambda =5200-6250$\AA\, before calculating a median stack.
The resulting stacked spectrum is shown in Fig.~\ref{fig:sp_stack}, along with
the best-fitting emission-line plus continuum model.

The measured value of EW(H$\alpha$+[N{\sc ii}]) from the stacked spectra is 
$189\pm 19$\AA. Assuming that H$\alpha$ contributes 90\% of the
H$\alpha$+[N{\sc ii}] flux \citep{Sanders2015}, the stacked spectrum therefore 
indicates a median EW(H$\alpha$) of $170\pm 17$\AA. In the inset, the distribution of 
EW(H$\alpha$) measurements from the SED fits to the 48 individual galaxies are 
shown as a histogram. The median value of the individual EW(H$\alpha$) 
determinations is $175\pm14$\AA, in excellent agreement with the value measured 
from the stacked spectra.

Although in the individual spectra it is not possible to reliably fit H$\beta$ and 
[S{\sc ii}]$\lambda\lambda$6717,6731 for most of the galaxies, these emission lines 
are clear in the stacked spectra. As can be seen from the results reported in 
Fig.~\ref{fig:sp_stack}, we find that, on average, $\simeq 12\%$ of the 
[S{\sc ii}]+H$\alpha$+[N{\sc ii}] flux is contributed by [S{\sc ii}].

\section{Results}\label{sec:results}
\begin{figure*}
\centering
\includegraphics[width=0.75\textwidth,clip=true]{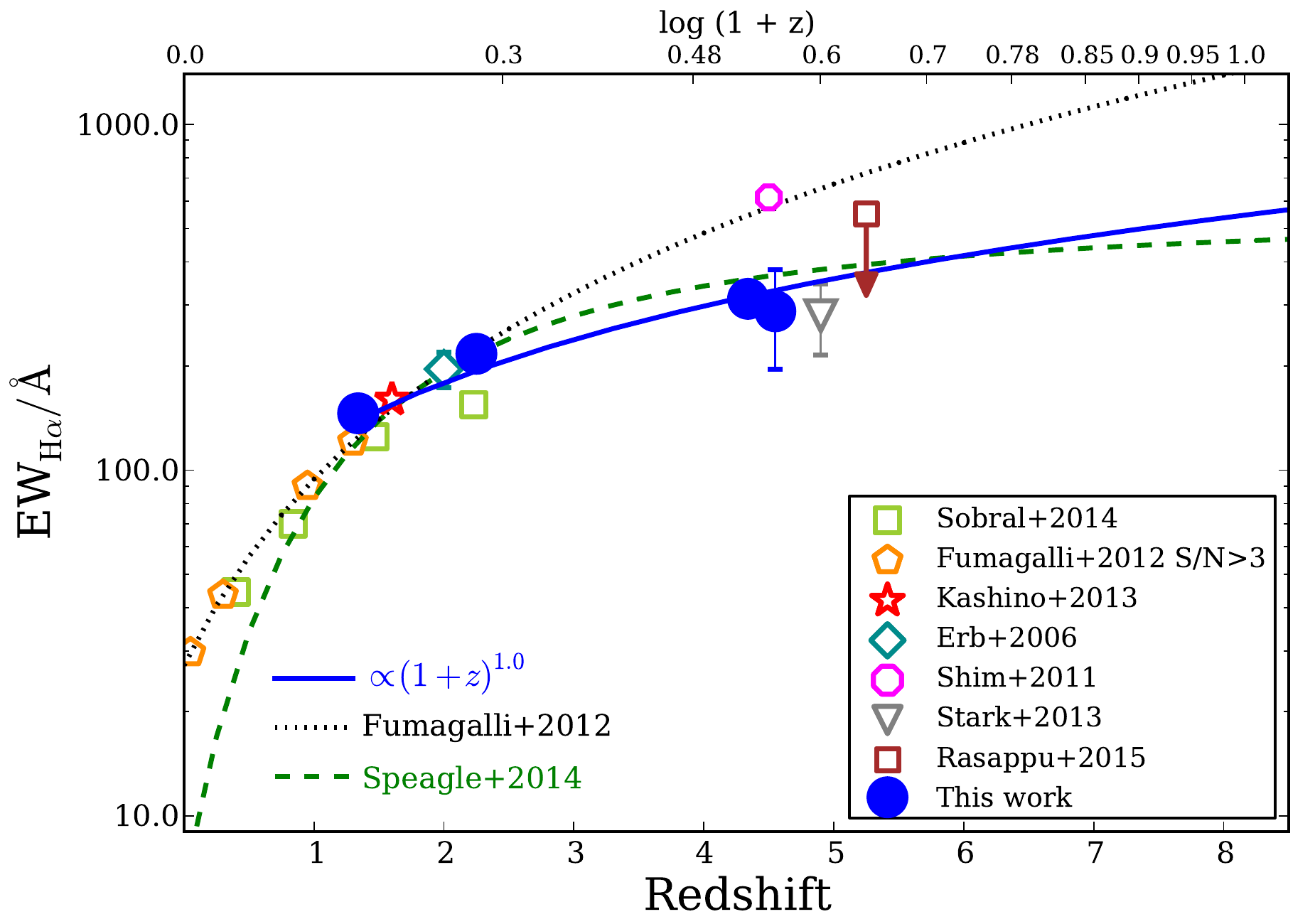}
\caption{The evolution of EW(H$\alpha$) as a function of redshift. 
The filled blue circles show our SED-fitting based EW(H$\alpha$) measurements (median values)
for the four different galaxy samples presented in Table~\ref{table_properties}. 
The results of various previous studies from the literature are also plotted for 
comparison. The grey dotted line shows the evolution extrapolation by 
\citet{Fumagalli2012}, based on galaxies at $z\leq 1.5$ with stellar masses in the 
range 10$^{10-10.5}$ and $S/N>3$ detections of the H$\alpha$ emission line
in their 3D-HST spectra. The green dashed line shows a fit to our EW(H$\alpha$) 
data assuming the same functional form as the sSFR$-z$ relation derived by 
\citet{Speagle2014} for a galaxy stellar mass of $\log(M_{\star}/\Msolar)=9.8$ 
(i.e. green dashed line in Fig.~\ref{fig:evolution_sSFR}). After adopting this 
fixed functional form, we have simply shifted the normalisation of the green dashed 
line to match our EW(H$\alpha$) data point at $z\simeq 2.3$. 
As in Fig.~\ref{fig:evolution_sSFR}, the blue solid line shows a relation of 
the form: $\propto (1+z)^{1.0}$, which was fit to our new data points alone 
(see text for discussion).}
\label{fig:evolution_EW}
\end{figure*}

We now present our new results on both sSFR and EW(H$\alpha$) and compare them with 
recent results in the literature. The basic results are tabulated in 
Table~\ref{table_properties} and plotted in Figs.~\ref{fig:evolution_sSFR} 
\& \ref{fig:evolution_EW}.

\subsection{The evolution of sSFR}\label{sec:sSFR_vs_z}
Before discussing our new EW(H$\alpha$) results, it is of interest to explore our 
determinations of sSFR within the context of recent literature studies.
This serves two purposes. Firstly, it is an opportunity to confirm
that our adopted methods for estimating sSFR are at least in
reasonable agreement with previous studies, and that our four galaxy
samples are representative of main-sequence galaxies within the
redshift interval of interest. Secondly, if there is a close link
between the evolution of sSFR and EW(H$\alpha$), the apparent evolution
of sSFR with redshift should provide a template for the redshift
evolution of EW(H$\alpha$).

In Fig.~\ref{fig:evolution_sSFR} we plot median sSFR against redshift for our four 
galaxy samples. In each case we plot two values of sSFR, one where the SFR is based on 
the dust-corrected UV luminosities and one where SFR is based on the H$\alpha$ line 
fluxes corresponding to the EW(H$\alpha$) measurements. In 
Fig.~\ref{fig:evolution_sSFR} we have also plotted the results from a selection of 
sSFR studies in the literature for comparison. To provide a fair
comparison, for those studies at $z\leq 4$ we have corrected the sSFR
values to a common stellar mass of $\log_{10}(M_{\star}/\Msolar)=9.8$
(Chabrier IMF) based on the prescription of \citet{Dutton2010}. 
For those literature studies at $z\geq 4$ the median
masses of the galaxy samples should be sufficiently close to
$\log_{10}(M_{\star}/\Msolar)=9.8$ to make any correction
unnecessary. The exception to this is the sSFR data point from \citet{Rasappu2015}, 
which we have plotted as a down arrow simply because it
is based on a galaxy sample with a significantly lower median stellar mass 
(see Section 5.3.3). Finally, in Fig.~\ref{fig:evolution_sSFR} we also plot three 
different curves. The first (green dashed line) is the sSFR$-z$ relation for galaxies 
with stellar mass of $\log_{10}(M_{\star}/\Msolar)=9.8$ from the meta-analysis of 25
different literature main-sequence studies by \citet{Speagle2014}. The second 
(black dotted line) is the theoretical
expectation that sSFR(z) $\propto (1+z)^{2.25}$, normalised to pass through the
cluster of sSFR data-points within the redshift range $1<z<2$. The
final curve (blue solid line) is the best-fitting sSFR$-z$ relationship
of the form: $\propto  (1+z)^{1.0}$, as derived by fitting to our new
data-points alone, and is therefore valid at $z > 1$. This sSFR$-z$ 
relationship is suggested by considering the evolution of EW(H$\alpha$) and sSFR 
together, and is discussed further in Section 6.

Two points are immediately clear from Fig.~\ref{fig:evolution_sSFR}. 
Firstly, it can be seen that our sSFR estimates based on EW(H$\alpha$)
are very consistent with the corresponding sSFR estimates based on UV
luminosity. Secondly, it can be seen that both sSFR determinations are
perfectly consistent with the sSFR$-z$ relation for galaxies with stellar mass of $\log_{10}(M_{\star}/\Msolar)=9.8$ from 
\citet{Speagle2014}, which also provides a good description of the
literature sSFR data over the full $0<z<7$ redshift range. We can
therefore be confident that our galaxy samples are representative of
typical star-forming galaxies within the interval $1<z<5$.
Moreover, circumstantially, the agreement between the EW(H$\alpha$)
and UV-based sSFR estimates suggests that EW(H$\alpha$) could
be a reasonable proxy for sSFR at $z\leq5$, an issue
which will be pursued further below. 

\subsection{The evolution of EW(H$\bmath \alpha$)}\label{sec:EW_vs_z}
In Fig.~\ref{fig:evolution_EW} we plot the median EW(H$\alpha$) for our four 
galaxy samples
against redshift, together with recent determinations of EW(H$\alpha$)
from the literature. In particular, at $z\leq2.5$ we plot the spectroscopic results 
of \citet{Fumagalli2012}, \citet{Erb2006} and \citet{Kashino2013}, 
and the narrow-band results of \citet{Sobral2014}. In order to ensure a fair 
comparison, where
possible, we have plotted the literature results appropriate for the
same stellar mass range as our own data 
(i.e. $9.5<\log_{10}{M_{\star}/\Msolar}<10.5$).
Where this is not possible, we have scaled the published EW(H$\alpha$) values under 
the assumption that EW(H$\alpha$) $\propto M_{\star}^{-0.25}$ as determined by 
\citet{Sobral2014}\footnote{Note:
  the EW(H$\alpha$) results of \cite{Erb2006}, \cite{Fumagalli2012}, and
  \cite{Kashino2013} are all consistent with a $M_{\star}^{-0.25}$ scaling.}.
Where necessary, we have converted quoted values of
EW(H$\alpha$+N[{\sc ii}]) to EW(H$\alpha$) by assuming that
EW(H$\alpha$)$=0.9\times$EW(H$\alpha$+N[{\sc ii}]), as recently derived by
\citet{Sanders2015} from near-IR spectroscopy.

At higher redshifts (z $\geq 4$) we plot the EW(H$\alpha$) results
from three previous studies (Shim et al. 2011; Stark et al 2013 and
Rasappu et al. 2015). As before, we plot the \citet{Rasappu2015}
data point with a down-arrow, simply because it is based on a sample
of galaxies with a significantly lower median stellar mass (see Section 5.3.3).
As in Fig.~\ref{fig:evolution_sSFR}, the blue solid line shows the best-fitting
EW(H$\alpha$)$-z$ relation of the form: $\propto (1+z)^{1.0}$, with a
normalisation set by fitting to our new data-points in the range $1<z<5$ 
alone. Likewise, the green dashed line shows an EW(H$\alpha$)$-z$ relation
with the same functional form as the \citet{Speagle2014} sSFR$-z$
relation, with the appropriate normalisation to pass through our new
data point at $z\simeq 2.3$. Finally, the black dotted line shows an
EW(H$\alpha$)$-z$ relation of the form: $\propto (1+z)^{1.8}$, as
derived by \citet{Fumagalli2012} at $z\leq 1.5$.

\subsection{Comparison with previous results}
Figure~\ref{fig:evolution_EW} demonstrates that our new measurements of 
EW(H$\alpha$) at $z\simeq 1.3$ and 
$z\simeq 2.3$ are in good agreement with previous literature results based on 
near-IR spectroscopy \citep[e.g.][]{Erb2006,Fumagalli2012,Kashino2013} and 
narrow-band imaging \citep[e.g.][]{Sobral2014}. It can be seen that at 
$0\leq z\leq 2$ a relatively consistent picture emerges, whereby the typical 
EW(H$\alpha$) displayed by $\simeq 10^{10}\Msolar$ star-forming galaxies evolves 
by a factor of $\simeq 7$, reaching a value of $\simeq 200$\AA\, by $z\simeq 2$. 
In contrast, several recent studies of the evolution of EW(H$\alpha$) at $z\geq 2$ 
have reached apparently contradictory conclusions, with quoted values for the 
typical EW(H$\alpha$) at $z\geq 4$ differing by factors of $\simeq 3$. 
In this sub-section we compare our new results to those of recent
high-redshift EW(H$\alpha$) studies and attempt to identify the
sources of any discrepancies.

\subsubsection{Shim et al. (2011)}
\citet{Shim2011} performed a very similar SED-based analysis to the work presented 
here, and derived EW(H$\alpha$) values for a final sample of 64 
spectroscopically-confirmed LBGs in the redshift range $3.8<z<5.0$, selected from 
GOODS-S and GOODS-N fields. In Fig.~\ref{fig:evolution_EW} we have plotted a 
rest-frame EW(H$\alpha$) of 615\AA\, for \citet{Shim2011}, which is the median value for 
the objects in common with our work, and after applying our estimated correction factor $f=0.79$ 
rather than the value of 0.71 quoted in their paper.
This number is a full factor of two larger than our new EW(H$\alpha$) result of 
$\simeq 300$\AA, which clearly requires some explanation. Although, \citet{Shim2011}
adopted different SED models (i.e. Charlot \& Bruzual 2007 models 
rather than BC03) and considered stellar populations with a lower age limit 
(i.e. 1 Myr rather than 50 Myr), our tests suggest that straightforward differences 
in photometry are responsible for the majority of the difference in our final 
EW(H$\alpha$) numbers.

A direct comparison of the photometry for the 19 GOODS-S objects which are common 
to \citet{Shim2011} and our own $z\simeq 4.5$ spectroscopic galaxy sample indicates 
that the \citet{Shim2011} IRAC 3.6$\mu$m photometry is systematically $\sim 0.2$ 
magnitudes brighter than the deconfused photometry from \citet{Guo2013}. If we 
artificially brightened our IRAC 3.6$\mu$m photometry by 0.2 magnitudes, this would 
be sufficient to raise our derived EW(H$\alpha$) value to $\simeq 590$\AA, in 
good agreement with the \citet{Shim2011} result. Any remaining difference can 
likely be attributed to the improved quality of the $K_{s}-$band photometry 
available within the CANDELS UDS and GOODS-S fields from the HUGS survey. Within 
this context, given that our $K_{s}-$band and IRAC photometry should be 
aperture matched by the {\sc tfit} algorithm based on the same $H_{160}$
priors, it seems likely that the EW(H$\alpha$) values derived by \citet{Shim2011}
could be overestimated by a factor of $\simeq 2$.

\subsubsection{Stark et al. (2013)}
\citet{Stark2013} performed an SED-based analysis of 45 spectroscopically-confirmed 
LBGs drawn from GOODS-S and GOODS-N in the redshift interval $3.8<z<5.0$ (30 objects 
in common with Shim et al. 2011). The SED fitting performed by \citet{Stark2013} is 
even more similar to that performed here, although restricted to 1/5 solar
metallicity and allowing ages as low as 5 Myr, and they estimated that the
H$\alpha$ emission line contributed 76\% of the observed EW. The final value derived 
by Stark et al. is EW(H$\alpha$) $= 270$\AA, in good agreement with our new
results. Interestingly, the value of 270\AA\, is derived from their SED fits which 
{\it include} the IRAC 3.6$\mu$m photometry, rather than {\it exclude} it. 
The figure for EW(H$\alpha$) derived by \citet{Stark2013} based on SED 
fits which exclude the IRAC 3.6$\mu$m photometry is actually 410\AA, $\simeq 40$\% higher 
than our new number.

Although it is difficult to be definitive, it seems likely that differences in the 
available $K_{s}-$band photometry are at least partly responsible. From our final 
sample of 26 spectroscopically confirmed $3.8<z<5.0$ galaxies, 15 are in common 
with the \citet{Stark2013} sample. For this sub-sample we find that the new HUGS 
$K_{s}-$band photometry is systematically $\simeq 0.15$ magnitudes brighter than 
the ISAAC $K_{s}-$band photometry available to \citet{Stark2013}. Assuming similar 
IRAC photometry, this difference alone is likely to explain most of the off-set 
between our new results and those of \citet{Stark2013}.

\subsubsection{Rasappu et al. (2015)}
Most recently, \citet{Rasappu2015} presented a study of EW(H$\alpha$+[NII]+[SII]) 
for a sample of star-forming galaxies at $5.1<z<5.4$ in the GOODS-N and GOODS-S 
fields. Based on an SED-fitting analysis (excluding the H$\alpha$ contaminated
IRAC 4.5$\mu$m filter) of 13 galaxies with spectroscopic redshifts and 11
galaxies with photometric redshifts in this redshift interval, Rasappu
et al. find mean EW(H$\alpha$+[NII]+[SII]) values of $665 \pm
53$\AA\, and $707 \pm 74$ \AA\, for the photometric and the spectroscopic samples 
respectively. If we simply average these values and assume that H$\alpha$ 
contributes $\simeq 80\%$ of the total EW(H$\alpha$+[NII]+[SII]), this
suggests EW(H$\alpha) \simeq 550$\AA. Alternatively, the median of the
individual EW(H$\alpha$+[NII]+[SII]) determinations from
\citet{Rasappu2015} is 670\AA, which again leads to EW(H$\alpha) \simeq 550$\AA.
It is therefore clear that the EW(H$\alpha$) results of
\citet{Rasappu2015} at $z\simeq5.3$ are a factor of $\simeq 1.8$
higher than our results at $z\simeq 4.5$.

Although apparently inconsistent, it is likely that most of this
discrepancy can be attributed to differences in the stellar masses of
the galaxies studied here and in \citet{Rasappu2015}.
After taking into account differences in the assumed IMF, only 3 galaxies in the 
Rasappu sample have stellar masses within our adopted range. The median
EW(H$\alpha$+[NII]+[SII]) for those three galaxies is 310$\AA$, implying EW(H$\alpha$)$\simeq250$\AA, in good agreement with our results.
In fact, the median
mass for the galaxies in their sample is $\log
({M_{\star}}/{\rm M}_{\odot}) \simeq 9.1$, (Salpeter IMF), $\simeq 0.9$ dex lower than 
the median mass of our $z\simeq 4.5$ galaxy samples (after converting
to a Chabrier IMF). If EW(H$\alpha$)
continues to scale as $\propto M_{\star}^{-0.25}$, this alone is enough to
account for a factor of $\simeq 1.7$ difference in the expected
EW(H$\alpha$). As discussed previously, for this reason we have
plotted the \citet{Rasappu2015} data point with a down arrow.

\section{The joint evolution of specific SFR and H$\bmath \alpha$ equivalent 
width}
In the introduction we highlighted that one of the reasons for
studying the evolution of EW(H$\alpha$) was the possibility that it 
could provide a useful proxy for sSFR at high redshift.
Within this context, it is worth remembering that, under the assumption 
that the relationship between H$\alpha$ line flux
and SFR remains fixed, the sSFR$-$EW(H$\alpha$) relationship might be
expected to evolve in concert with the average $M/L$ ratios of main
sequence galaxies. Indeed, the results presented in Table 1 provide some evidence that
this is the case over the redshift interval $1<z<5$, which sees sSFR
(UV based) rise by a factor of $\simeq 2.7$ compared to a factor of 
$\simeq 2.1$ increase in EW(H$\alpha$) in the same redshift range.  
However, most of this difference occurs between $1<z<2$, with sSFR
and EW(H$\alpha$) both increasing by very similar amounts with redshift 
within the range $2<z<5$. This is consistent with our finding that the
average $M/L$ ratios of main sequence galaxies with stellar masses of
$\simeq 10^{10}\Msolar$ change very little between $z\simeq 2.3$ and
$z\simeq 4.5$. Moreover, from an empirical perspective, an inspection of the
results presented in Figs.~\ref{fig:evolution_sSFR} \&
\ref{fig:evolution_EW} suggests that the redshift evolution displayed by sSFR and EW(H$\alpha$) is very similar. 
Motivated by this, in this section we employ two different methods to investigate the
joint evolution of sSFR and EW(H$\alpha$).

Firstly, we rely on our new EW(H$\alpha$) and sSFR data points at $1<z<5$ alone. 
Our data points in Figs.~\ref{fig:evolution_sSFR} \& \ref{fig:evolution_EW} 
immediately suggest evolution which should be adequately described by a simply
power-law of the form: $\propto (1+z)^{n}$. Indeed, if we fit a power-law of
this form to our data points alone, we find best-fitting relations of the form:
\begin{equation}
sSFR(z) \propto (1+z)^{1.2\pm0.4}
\end{equation}
\begin{equation}
EW(z) \propto (1+z)^{0.8\pm 0.2}
\end{equation}
\noindent
for the evolution of sSFR and EW(H$\alpha$) respectively. As a result,
it would seem reasonable to fit the evolution of both quantities using
a power-law of the form: $(1+z)^{1.0}$; which is shown as the blue 
solid lines in Figs.~\ref{fig:evolution_sSFR} \& \ref{fig:evolution_EW}. 
Clearly, the simple ratio of these two expressions 
provides us with a conversion between sSFR and EW(H$\alpha$).
This leads to a relationship of the form: 
\begin{equation}
EW(H\alpha)=(59\pm10)\times sSFR
\end{equation}
\noindent
where EW(H$\alpha$) is measured in \AA\, and sSFR is measured in
Gyr$^{-1}$. It can be seen from Figs.~\ref{fig:evolution_sSFR} \& \ref{fig:evolution_EW} 
that a relationship of this form provides a reasonably good
description of the evolution of both EW(H$\alpha$) and sSFR over the redshift 
interval $1<z<5$.

The second approach we adopt is based on the sSFR$-z$ relation from
the meta-analysis of \citet{Speagle2014}. In Fig.~\ref{fig:evolution_sSFR} the 
sSFR$-z$ relation from \citet{Speagle2014} for galaxies with stellar mass
$\log_{10}(M_{\star}/\Msolar)=9.8$ is shown as the dashed green line. It can be
seen that this functional form provides an excellent match to the
observational data over the full $0<z<7$ redshift range shown in the figure.
Under the assumption that sSFR and EW(H$\alpha$) display the same
evolution with redshift, we are free to use the same functional form
in Fig.~\ref{fig:evolution_EW}, choosing to floating the normalisation such that the 
Speagle et al. curve passes through our new data-point at $z\simeq 2.3$. 
Clearly this functional form actually does a reasonable job of describing
the evolution of EW(H$\alpha$) over the redshift range
$0.5<z<5.0$. Again, the ratio of the two dashed green curves shown in
Figs.~\ref{fig:evolution_sSFR} \& \ref{fig:evolution_EW} provide us with a 
conversion between sSFR and EW(H$\alpha$). 
In this case, the resulting relationship is of the form:
\begin{equation}
EW(H\alpha)=(68\pm10)\times sSFR
\end{equation}
\noindent
which can be seen to be in good agreement with the normalisation derived from our 
new data points alone. Based on these results, we would argue that EW(H$\alpha$) 
remains a useful independent tracer of sSFR over the redshift interval $1<z<5$, and 
that the conversion between the two quantities is approximately: 
\begin{equation}
EW(H\alpha)=(63\pm 7)\times sSFR
\end{equation}
\noindent

Our finding that EW(H$\alpha$) evolves relatively slowly over
the redshift interval $1<z<5$ is in contrast to at least some previous
results in the literature. However, as was discussed in Section 5.3,
many of the apparent discrepancies can either be explained by our
improved photometry, or the effects of trying to compare sSFR/EW(H$\alpha$)
evolution amongst samples with very different stellar masses. 
While the median EW(H$\alpha$) and sSFR do increase significantly from $z\simeq1$ 
to $z\simeq2$, their evolution from $z\simeq 2$ to $z\simeq 5$ is noticeable less dramatic.

Although measuring EW(H$\alpha$) via SED-fitting is not trivial,
it does possess the distinct advantage of being largely independent of
the age, nebular emission and dust uncertainties which plague SED-based
measurements of star-formation rate. Fortunately, it will soon be relatively trivial to directly {\it measure} 
the EW(H$\alpha$) for star-forming galaxies in the
redshift interval $0.5<z<6.5$ with NirSpec on JWST. If the evolution
follows a steep trend with redshift, such as an extrapolation of the
\citet{Fumagalli2012} results (dotted line in Fig. ~\ref{fig:evolution_EW}), 
then star-forming galaxies 
at $z\simeq 6.5$ with stellar masses of $\simeq
10^{10}\Msolar$ should have EW(H$\alpha) \simeq 1000$\AA. In
contrast, if the evolution of EW(H$\alpha$) follows the shallower trend
suggested by our new results, the same galaxies should have
EW(H$\alpha) \simeq 450$\AA.

\section{Summary}\label{sec:summary}

In this paper we have presented the results of a study aimed at improving our
understanding of how EW(H$\alpha$) evolves with redshift and testing whether or not 
EW(H$\alpha$) can be exploited as an independent proxy for sSFR at high redshift.

Based on a sample of star-forming galaxies at $1.2<z<1.5$, where is is
possible to directly measure EW(H$\alpha$) from 3D-HST grism spectroscopy,
we first demonstrated that it is possible to reliably measure
EW(H$\alpha$) via SED-fitting of the multi-wavelength photometry
available in the CANDELS UDS and GOODS-S fields.

Having demonstrated the validity of our technique, we then proceeded to
explore the redshift evolution of EW(H$\alpha$), using samples of
spectroscopically confirmed galaxies at $1.2<z<1.5$, $2.1<z<2.45$
and $3.8<z<5.0$ for which the H$\alpha$ emission line contaminates 
the $H_{160}, K_{s}$ and IRAC 3.6$\mu$m filters respectively. To improve our
high-redshift statistics, we also measured EW(H$\alpha$) in a
photometric-redshift selected sample of $3.8<z<5.0$ star-forming
galaxies, finding excellent agreement with the results derived from
the spectroscopically confirmed sample. In order to ameliorate
potential trends with stellar mass, all of our galaxy samples are
restricted to the mass range $9.5< \log(M_{\star}/\Msolar) < 10.5$  and have
median masses in the range $9.7< \log(M_{\star}/\Msolar)<9.8$.

Combining stellar-mass estimates derived from SED fitting with
measurements of UV luminosity and EW(H$\alpha$), we derived two
measurements of sSFR for each of our four galaxy samples. Both
measurements are found to be consistent, and in excellent agreement
with recent determinations of the evolution of the so-called main
sequence of star-formation. Having demonstrated that our galaxies are fully 
consistent with being located on the main
sequence, we compared our new results for the evolution of
EW(H$\alpha$) with recent determinations in the literature. We
concluded that many of the apparent discrepancies with previous
literature results are either caused by simple differences in
photometry or by comparing the EW(H$\alpha$)/sSFR values of galaxy samples with
significantly different stellar masses.

Comparing the evolution of sSFR and EW(H$\alpha$), two
different methods were employed to demonstrate that sSFR and EW(H$\alpha$) are
consistent with displaying the same evolution with redshift.
Taken together, our new results suggest that EW(H$\alpha$)/sSFR evolve
relatively slowly ($\propto (1+z)^{1.0}$), increasing with redshift 
by a factor of $\leq 3$ over the redshift interval $1<z<5$. 
We conclude that over the interval $1<z<5$, 
EW(H$\alpha$) can serve as a useful independent proxy for sSFR, and that the relative
normalisation of the two quantities is: EW(H$\alpha$)$ =(63\pm7)\times$ sSFR.
If correct, this form of evolution would predict that by $z\simeq
6.5$, galaxies with stellar masses $\simeq 10^{10}\Msolar$ will
display an average EW(H$\alpha$) of $\simeq 450\AA$\, and have sSFR values of 
$\simeq 6.5$ Gyr$^{-1}$. Fortunately, it will be possible to obtain accurate 
measurements of both quantities following the launch of JWST.

\section*{Acknowledgements}

EMQ would like to thank to the high-z group at Edinburgh, and to Emma Curtis-Lake in
particular, for very interesting, supportive and constructive talks.
The authors are grateful to Mattia Fumagalli and David Sobral for providing their 
published results for a direct comparison. We also thank the referee for
suggestions which helped to clarify the text. EMQ and RJM acknowledge the 
support of the European Research Council via the award of a Consolidator Grant 
(PI McLure), 
while FC acknowledges the support of the Science and Technology Facilities Council 
(STFC) via the award of an STFC Studentship. JSD acknowledges the support of the 
European Research Council through the award of an Advanced Grant. The research 
leading to these results has received funding from the European Union 
Seventh Framework Programme (FP7/2007-2013) under grant agreement n$^{\circ}$ 312725 
(ASTRODEEP).
This work uses data from the following ESO programs: 181.A0717, 186.A-0898.
This work is based on observations taken by the CANDELS Multi-Cycle Treasury 
Program and the 3D-HST Treasury Program (GO 12177 and 12328) with the NASA/ESA 
HST, which is operated by the Association of Universities for Research in 
Astronomy, Inc., under NASA contract NAS5-26555.

\bibliography{biblio_highz}

\begin{thebibliography}{59}
\expandafter\ifx\csname natexlab\endcsname\relax\def\natexlab#1{#1}\fi

\bibitem[{{Brammer} {et~al}\mbox{.}(2012){Brammer}, {van Dokkum}, {Franx},
  {Fumagalli}, \& {et al.}}]{Brammer2012}
{Brammer} G.~B., {van Dokkum} P.~G., {Franx} M., {Fumagalli} M., {et al.},
  2012, ApJS, 200, 13

\bibitem[{{Bruzual} \& {Charlot}(2003)}]{BC03}
{Bruzual} G., {Charlot} S., 2003, MNRAS, 344, 1000

\bibitem[{{Calzetti} {et~al}\mbox{.}(2000){Calzetti}, {Armus}, {Bohlin},
  {Kinney}, {Koornneef}, \& {Storchi-Bergmann}}]{Calzetti2000}
{Calzetti} D., {Armus} L., {Bohlin} R.~C., {Kinney} A.~L., {Koornneef} J.,
  {Storchi-Bergmann} T., 2000, ApJ, 533, 682

\bibitem[{{Calzetti}, {Kinney} \& {Storchi-Bergmann}(1994){Calzetti}, {Kinney},
  \& {Storchi-Bergmann}}]{Calzetti1994}
{Calzetti} D., {Kinney} A.~L., {Storchi-Bergmann} T., 1994, ApJ, 429, 582

\bibitem[{{Cimatti} {et~al}\mbox{.}(2002){Cimatti}, {Mignoli}, {Daddi},
  {Pozzetti}, {Fontana}, {Saracco}, {Poli}, {Renzini}, {Zamorani},
  {Broadhurst}, {Cristiani}, {D'Odorico}, {Giallongo}, {Gilmozzi}, \&
  {Menci}}]{Cimatti2002}
{Cimatti} A. {et~al.}, 2002, A\&A, 392, 395

\bibitem[{{Cooper} {et~al}\mbox{.}(2012){Cooper}, {Yan}, {Dickinson}, {Juneau},
  {Lotz}, {Newman}, {Papovich}, {Salim}, {Walth}, {Weiner}, \&
  {Willmer}}]{Cooper2012}
{Cooper} M.~C. {et~al.}, 2012, MNRAS, 425, 2116

\bibitem[{{Cullen} {et~al}\mbox{.}(2014){Cullen}, {Cirasuolo}, {McLure},
  {Dunlop}, \& {Bowler}}]{Cullen2014}
{Cullen} F., {Cirasuolo} M., {McLure} R.~J., {Dunlop} J.~S., {Bowler} R.~A.~A.,
  2014, MNRAS, 440, 2300

\bibitem[{{Curtis-Lake} {et~al}\mbox{.}(2013){Curtis-Lake}, {McLure}, {Dunlop},
  {Schenker}, {Rogers}, {Targett}, {Cirasuolo}, {Almaini}, {Ashby}, {Bradshaw},
  {Finkelstein}, {Dickinson}, {Ellis}, {Faber}, {Fazio}, {Ferguson}, {Fontana},
  {Grogin}, {Hartley}, {Kocevski}, {Koekemoer}, {Lai}, {Robertson}, {Vanzella},
  \& {Willner}}]{Curtis-Lake2013}
{Curtis-Lake} E. {et~al.}, 2013, MNRAS, 429, 302

\bibitem[{{Daddi} {et~al}\mbox{.}(2007){Daddi}, {Dickinson}, {Morrison},
  {Chary}, {Cimatti}, {Elbaz}, {Frayer}, {Renzini}, {Pope}, {Alexander},
  {Bauer}, {Giavalisco}, {Huynh}, {Kurk}, \& {Mignoli}}]{Daddi2007}
{Daddi} E. {et~al.}, 2007, ApJ, 670, 156

\bibitem[{{Dav{\'e}}, {Oppenheimer} \& {Finlator}(2011){Dav{\'e}},
  {Oppenheimer}, \& {Finlator}}]{Dave2011}
{Dav{\'e}} R., {Oppenheimer} B.~D., {Finlator} K., 2011, MNRAS, 415, 11

\bibitem[{{de Barros}, {Schaerer} \& {Stark}(2014){de Barros}, {Schaerer}, \&
  {Stark}}]{deBarros2014}
{de Barros} S., {Schaerer} D., {Stark} D.~P., 2014, A\&A, 563, A81

\bibitem[{{Dutton}, {van den Bosch} \& {Dekel}(2010){Dutton}, {van den Bosch},
  \& {Dekel}}]{Dutton2010}
{Dutton} A.~A., {van den Bosch} F.~C., {Dekel} A., 2010, MNRAS, 405, 1690

\bibitem[{{Elbaz} {et~al}\mbox{.}(2007){Elbaz}, {Daddi}, {Le Borgne},
  {Dickinson}, {Alexander}, {Chary}, {Starck}, {Brandt}, {Kitzbichler},
  {MacDonald}, {Nonino}, {Popesso}, {Stern}, \& {Vanzella}}]{Elbaz2007}
{Elbaz} D. {et~al.}, 2007, A\&A, 468, 33

\bibitem[{{Erb} {et~al}\mbox{.}(2006){Erb}, {Steidel}, {Shapley}, {Pettini},
  {Reddy}, \& {Adelberger}}]{Erb2006}
{Erb} D.~K., {Steidel} C.~C., {Shapley} A.~E., {Pettini} M., {Reddy} N.~A.,
  {Adelberger} K.~L., 2006, ApJ, 647, 128

\bibitem[{{Fontana} {et~al}\mbox{.}(2014){Fontana}, {Dunlop}, {Paris},
  {Targett}, {Boutsia}, {Castellano}, {Galametz}, {Grazian}, {McLure},
  {Merlin}, {Pentericci}, {Wuyts}, {Almaini}, {Caputi}, {Chary}, {Cirasuolo},
  {Conselice}, {Cooray}, {Daddi}, {Dickinson}, {Faber}, {Fazio}, {Ferguson},
  {Giallongo}, {Giavalisco}, {Grogin}, {Hathi}, {Koekemoer}, {Koo}, {Lucas},
  {Nonino}, {Rix}, {Renzini}, {Rosario}, {Santini}, {Scarlata}, {Sommariva},
  {Stark}, {van der Wel}, {Vanzella}, {Wild}, {Yan}, \& {Zibetti}}]{HUGS}
{Fontana} A. {et~al.}, 2014, A\&A, 570, A11

\bibitem[{{Fumagalli} {et~al}\mbox{.}(2012){Fumagalli}, {Patel}, {Franx},
  {Brammer}, {van Dokkum}, {da Cunha}, {Kriek}, {Lundgren}, {Momcheva}, {Rix},
  {Schmidt}, {Skelton}, {Whitaker}, {Labbe}, \& {Nelson}}]{Fumagalli2012}
{Fumagalli} M. {et~al.}, 2012, ApJL, 757, L22

\bibitem[{{Galametz} {et~al}\mbox{.}(2013){Galametz}, {Grazian}, {Fontana},
  {Ferguson}, \& {the CANDELS Team}}]{Galametz2013}
{Galametz} A., {Grazian} A., {Fontana} A., {Ferguson} H.~C., {the CANDELS
  Team}, 2013, ApJS, 206, 10

\bibitem[{{Gonz{\'a}lez} {et~al}\mbox{.}(2014){Gonz{\'a}lez}, {Bouwens},
  {Illingworth}, {Labb{\'e}}, {Oesch}, {Franx}, \& {Magee}}]{Gonzalez2014}
{Gonz{\'a}lez} V., {Bouwens} R., {Illingworth} G., {Labb{\'e}} I., {Oesch} P.,
  {Franx} M., {Magee} D., 2014, ApJ, 781, 34

\bibitem[{{Gonz{\'a}lez} {et~al}\mbox{.}(2010){Gonz{\'a}lez}, {Labb{\'e}},
  {Bouwens}, {Illingworth}, {Franx}, {Kriek}, \& {Brammer}}]{Gonzalez2010}
{Gonz{\'a}lez} V., {Labb{\'e}} I., {Bouwens} R.~J., {Illingworth} G., {Franx}
  M., {Kriek} M., {Brammer} G.~B., 2010, ApJ, 713, 115

\bibitem[{{Grogin} {et~al}\mbox{.}(2011){Grogin}, {Kocevski}, {Faber}, \& {the
  CANDELS Team}}]{Grogin2011}
{Grogin} N.~A., {Kocevski} D.~D., {Faber} S.~M., {the CANDELS Team}, 2011,
  ApJS, 197, 35

\bibitem[{{Guo} {et~al}\mbox{.}(2013){Guo}, {Ferguson}, {Giavalisco}, {Barro},
  \& {the CANDELS Team}}]{Guo2013}
{Guo} Y., {Ferguson} H.~C., {Giavalisco} M., {Barro}, {the CANDELS Team}, 2013,
  ApJS, 207, 24

\bibitem[{{Johnston} {et~al}\mbox{.}(2015){Johnston}, {Vaccari}, {Jarvis},
  {Smith}, {Giovannoli}, {H{\"a}u{\ss}ler}, \& {Prescott}}]{Johnston2015}
{Johnston} R., {Vaccari} M., {Jarvis} M., {Smith} M., {Giovannoli} E.,
  {H{\"a}u{\ss}ler} B., {Prescott} M., 2015, MNRAS, 453, 2540

\bibitem[{{Karim} {et~al}\mbox{.}(2011){Karim}, {Schinnerer},
  {Mart{\'{\i}}nez-Sansigre}, {Sargent}, {van der Wel}, {Rix}, {Ilbert},
  {Smol{\v c}i{\'c}}, {Carilli}, {Pannella}, {Koekemoer}, {Bell}, \&
  {Salvato}}]{Karim2011}
{Karim} A. {et~al.}, 2011, ApJ, 730, 61

\bibitem[{{Kashino} {et~al}\mbox{.}(2013){Kashino}, {Silverman}, {Rodighiero},
  {Renzini}, {Arimoto}, {Daddi}, {Lilly}, {Sanders}, {Kartaltepe}, {Zahid},
  {Nagao}, {Sugiyama}, {Capak}, {Carollo}, {Chu}, {Hasinger}, {Ilbert},
  {Kajisawa}, {Kewley}, {Koekemoer}, {Kova{\v c}}, {Le F{\`e}vre}, {Masters},
  {McCracken}, {Onodera}, {Scoville}, {Strazzullo}, {Symeonidis}, \&
  {Taniguchi}}]{Kashino2013}
{Kashino} D. {et~al.}, 2013, ApJL, 777, L8

\bibitem[{{Kennicutt} \& {Evans}(2012)}]{Kennicutt-Evans2012}
{Kennicutt} R.~C., {Evans} N.~J., 2012, ARA\&A, 50, 531

\bibitem[{{Koekemoer} {et~al}\mbox{.}(2011){Koekemoer}, {Faber}, {Ferguson}, \&
  {Grogin}}]{Koekemoer2011}
{Koekemoer} A.~M., {Faber} S.~M., {Ferguson} H.~C., {Grogin} t., 2011, ApJS,
  197, 36

\bibitem[{{Koprowski} {et~al}\mbox{.}(2015){Koprowski}, {Dunlop},
  {Michalowski}, {Roseboom}, {Geach}, {Cirasuolo}, {Aretxaga}, {Bowler},
  {Banerji}, {Bourne}, {Coppin}, {Chapman}, {Hughes}, {Jenness}, {McLure},
  {Symeonidis}, \& {van der Werf}}]{Koprowski2015}
{Koprowski} M. {et~al.}, 2015, ArXiv e-prints

\bibitem[{{K{\"u}mmel} {et~al}\mbox{.}(2009){K{\"u}mmel}, {Walsh}, {Pirzkal},
  {Kuntschner}, \& {Pasquali}}]{Kummel2009}
{K{\"u}mmel} M., {Walsh} J.~R., {Pirzkal} N., {Kuntschner} H., {Pasquali} A.,
  2009, PASP, 121, 59

\bibitem[{{Laidler} {et~al}\mbox{.}(2007){Laidler}, {Papovich}, {Grogin},
  {Idzi}, {Dickinson}, {Ferguson}, {Hilbert}, {Clubb}, \&
  {Ravindranath}}]{Laidler2007}
{Laidler} V.~G. {et~al.}, 2007, PASP, 119, 1325

\bibitem[{{Le F{\`e}vre} {et~al}\mbox{.}(2004){Le F{\`e}vre}, {Vettolani},
  {Paltani}, {Tresse}, {Zamorani}, {Le Brun}, {Moreau}, {Bottini}, {Maccagni},
  {Picat}, {Scaramella}, {Scodeggio}, {Zanichelli}, {Adami}, {Arnouts},
  {Bardelli}, {Bolzonella}, {Cappi}, {Charlot}, {Contini}, {Foucaud},
  {Franzetti}, {Garilli}, {Gavignaud}, {Guzzo}, {Ilbert}, {Iovino},
  {McCracken}, {Mancini}, {Marano}, {Marinoni}, {Mathez}, {Mazure}, {Meneux},
  {Merighi}, {Pell{\`o}}, {Pollo}, {Pozzetti}, {Radovich}, {Zucca},
  {Arnaboldi}, {Bondi}, {Bongiorno}, {Busarello}, {Ciliegi}, {Gregorini},
  {Mellier}, {Merluzzi}, {Ripepi}, \& {Rizzo}}]{VVDS}
{Le F{\`e}vre} O. {et~al.}, 2004, A\&A, 428, 1043

\bibitem[{{Madau}(1995)}]{Madau1995}
{Madau} P., 1995, ApJ, 441, 18

\bibitem[{{Madau} \& {Dickinson}(2014)}]{Madau-Dickinson2014}
{Madau} P., {Dickinson} M., 2014, ARA\&A, 52, 415

\bibitem[{{Madau}, {Pozzetti} \& {Dickinson}(1998){Madau}, {Pozzetti}, \&
  {Dickinson}}]{Madau1998}
{Madau} P., {Pozzetti} L., {Dickinson} M., 1998, ApJ, 498, 106

\bibitem[{{McLeod} {et~al}\mbox{.}(2015){McLeod}, {McLure}, {Dunlop},
  {Robertson}, {Ellis}, \& {Targett}}]{McLeod2015}
{McLeod} D.~J., {McLure} R.~J., {Dunlop} J.~S., {Robertson} B.~E., {Ellis}
  R.~S., {Targett} T.~A., 2015, MNRAS, 450, 3032

\bibitem[{{McLure} {et~al}\mbox{.}(2011){McLure}, {Dunlop}, {de Ravel},
  {Cirasuolo}, {Schenker}, {Robertson}, {Koekemoer}, {Stark}, \&
  {Bowler}}]{McLure2011}
{McLure} R.~J. {et~al.}, 2011, MNRAS, 418, 2074

\bibitem[{{McLure} {et~al}\mbox{.}(2013){McLure}, {Pearce}, {Dunlop},
  {Cirasuolo}, {Curtis-Lake}, {Bruce}, {Caputi}, {Almaini}, {Bonfield},
  {Bradshaw}, {Buitrago}, {Chuter}, {Foucaud}, {Hartley}, \&
  {Jarvis}}]{McLure2013}
{McLure} R.~J. {et~al.}, 2013, MNRAS, 428, 1088

\bibitem[{{Meurer}, {Heckman} \& {Calzetti}(1999){Meurer}, {Heckman}, \&
  {Calzetti}}]{Meurer1999}
{Meurer} G.~R., {Heckman} T.~M., {Calzetti} D., 1999, ApJ, 521, 64

\bibitem[{{Mignoli} {et~al}\mbox{.}(2005){Mignoli}, {Cimatti}, {Zamorani},
  {Pozzetti}, {Daddi}, {Renzini}, {Broadhurst}, {Cristiani}, {D'Odorico},
  {Fontana}, {Giallongo}, {Gilmozzi}, {Menci}, \& {Saracco}}]{Mignoli2005}
{Mignoli} M. {et~al.}, 2005, A\&A, 437, 883

\bibitem[{{Mobasher} {et~al}\mbox{.}(2015){Mobasher}, {Dahlen}, {Ferguson},
  {Acquaviva}, {Barro}, {Finkelstein}, {Fontana}, {Gruetzbauch}, {Johnson},
  {Lu}, {Papovich}, {Pforr}, {Salvato}, {Somerville}, {Wiklind}, {Wuyts},
  {Ashby}, {Bell}, {Conselice}, {Dickinson}, {Faber}, {Fazio}, {Finlator},
  {Galametz}, {Gawiser}, {Giavalisco}, {Grazian}, {Grogin}, {Guo}, {Hathi},
  {Kocevski}, {Koekemoer}, {Koo}, {Newman}, {Reddy}, {Santini}, \&
  {Wechsler}}]{Mobasher2015}
{Mobasher} B. {et~al.}, 2015, ApJ, 808, 101

\bibitem[{{Morris} {et~al}\mbox{.}(2015){Morris}, {Kocevski}, {Trump},
  {Weiner}, {Hathi}, {Barro}, {Dahlen}, {Faber}, {Finkelstein}, {Fontana},
  {Ferguson}, {Grogin}, {Gr{\"u}tzbauch}, {Guo}, {Hsu}, {Koekemoer}, {Koo},
  {Mobasher}, {Pforr}, {Salvato}, {Wiklind}, \& {Wuyts}}]{Morris2015}
{Morris} A.~M. {et~al.}, 2015, ArXiv e-prints

\bibitem[{{Neistein} \& {Dekel}(2008)}]{Neistein2008}
{Neistein} E., {Dekel} A., 2008, MNRAS, 383, 615

\bibitem[{{Noeske} {et~al}\mbox{.}(2007){Noeske}, {Weiner}, {Faber},
  {Papovich}, {Koo}, {Somerville}, {Bundy}, {Conselice}, {Newman},
  {Schiminovich}, {Le Floc'h}, {Coil}, {Rieke}, {Lotz}, {Primack}, {Barmby},
  {Cooper}, {Davis}, {Ellis}, {Fazio}, {Guhathakurta}, {Huang}, {Kassin},
  {Martin}, {Phillips}, {Rich}, {Small}, {Willmer}, \& {Wilson}}]{Noeske2007}
{Noeske} K.~G. {et~al.}, 2007, ApJL, 660, L43

\bibitem[{{Oke}(1974)}]{AB_system1}
{Oke} J.~B., 1974, ApJS, 27, 21

\bibitem[{{Oke} \& {Gunn}(1983)}]{AB_system2}
{Oke} J.~B., {Gunn} J.~E., 1983, ApJ, 266, 713

\bibitem[{{Popesso} {et~al}\mbox{.}(2009){Popesso}, {Dickinson}, {Nonino},
  {Vanzella}, {Daddi}, {Fosbury}, {Kuntschner}, {Mainieri}, {Cristiani},
  {Cesarsky}, {Giavalisco}, {Renzini}, \& {GOODS Team}}]{Popesso2009}
{Popesso} P. {et~al.}, 2009, A\&A, 494, 443

\bibitem[{{Rasappu} {et~al}\mbox{.}(2015){Rasappu}, {Smit}, {Labbe}, {Bouwens},
  {Stark}, {Ellis}, \& {Oesch}}]{Rasappu2015}
{Rasappu} N., {Smit} R., {Labbe} I., {Bouwens} R., {Stark} D., {Ellis} R.,
  {Oesch} P., 2015, ArXiv e-prints

\bibitem[{{Reddy} {et~al}\mbox{.}(2015){Reddy}, {Kriek}, {Shapley}, {Freeman},
  {Siana}, {Coil}, {Mobasher}, {Price}, {Sanders}, \& {Shivaei}}]{Reddy2015}
{Reddy} N.~A. {et~al.}, 2015, ApJ, 806, 259

\bibitem[{{Renzini} \& {Peng}(2015)}]{Renzini-Peng2015}
{Renzini} A., {Peng} Y.-j., 2015, ApJL, 801, L29

\bibitem[{{Sanders} {et~al}\mbox{.}(2015){Sanders}, {Shapley}, {Kriek},
  {Reddy}, {Freeman}, {Coil}, {Siana}, {Mobasher}, {Shivaei}, {Price}, \& {de
  Groot}}]{Sanders2015}
{Sanders} R.~L. {et~al.}, 2015, ApJ, 799, 138

\bibitem[{{Schaerer} \& {de Barros}(2009)}]{Schaerer2009}
{Schaerer} D., {de Barros} S., 2009, A\&A, 502, 423

\bibitem[{{Shim} {et~al}\mbox{.}(2011){Shim}, {Chary}, {Dickinson}, {Lin},
  {Spinrad}, {Stern}, \& {Yan}}]{Shim2011}
{Shim} H., {Chary} R.-R., {Dickinson} M., {Lin} L., {Spinrad} H., {Stern} D.,
  {Yan} C.-H., 2011, ApJ, 738, 69

\bibitem[{{Sobral} {et~al}\mbox{.}(2013){Sobral}, {Smail}, {Best}, {Geach},
  {Matsuda}, {Stott}, {Cirasuolo}, \& {Kurk}}]{Sobral2014}
{Sobral} D., {Smail} I., {Best} P.~N., {Geach} J.~E., {Matsuda} Y., {Stott}
  J.~P., {Cirasuolo} M., {Kurk} J., 2013, MNRAS, 428, 1128

\bibitem[{{Speagle} {et~al}\mbox{.}(2014){Speagle}, {Steinhardt}, {Capak}, \&
  {Silverman}}]{Speagle2014}
{Speagle} J.~S., {Steinhardt} C.~L., {Capak} P.~L., {Silverman} J.~D., 2014,
  ApJS, 214, 15

\bibitem[{{Stark} {et~al}\mbox{.}(2009){Stark}, {Ellis}, {Bunker}, {Bundy},
  {Targett}, {Benson}, \& {Lacy}}]{Stark2009}
{Stark} D.~P., {Ellis} R.~S., {Bunker} A., {Bundy} K., {Targett} T., {Benson}
  A., {Lacy} M., 2009, ApJ, 697, 1493

\bibitem[{{Stark} {et~al}\mbox{.}(2013){Stark}, {Schenker}, {Ellis},
  {Robertson}, {McLure}, \& {Dunlop}}]{Stark2013}
{Stark} D.~P., {Schenker} M.~A., {Ellis} R., {Robertson} B., {McLure} R.,
  {Dunlop} J., 2013, ApJ, 763, 129

\bibitem[{{Tasca} {et~al}\mbox{.}(2015){Tasca}, {Le F{\`e}vre}, {Hathi},
  {Schaerer}, {Ilbert}, {Zamorani}, {Lemaux}, {Cassata}, {Garilli}, {Le Brun},
  {Maccagni}, {Pentericci}, {Thomas}, {Vanzella}, {Zucca}, {Amorin},
  {Bardelli}, {Cassar{\`a}}, {Castellano}, {Cimatti}, {Cucciati}, {Durkalec},
  {Fontana}, {Giavalisco}, {Grazian}, {Paltani}, {Ribeiro}, {Scodeggio},
  {Sommariva}, {Talia}, {Tresse}, {Vergani}, {Capak}, {Charlot}, {Contini}, {de
  la Torre}, {Dunlop}, {Fotopoulou}, {Koekemoer}, {L{\'o}pez-Sanjuan},
  {Mellier}, {Pforr}, {Salvato}, {Scoville}, {Taniguchi}, \&
  {Wang}}]{Tasca2015}
{Tasca} L.~A.~M. {et~al.}, 2015, A\&A, 581, A54

\bibitem[{{Vanzella} {et~al}\mbox{.}(2008){Vanzella}, {Cristiani}, {Dickinson},
  {Giavalisco}, {Kuntschner}, {Haase}, {Nonino}, {Rosati}, {Cesarsky},
  {Ferguson}, {Fosbury}, {Grazian}, {Moustakas}, {Rettura}, {Popesso},
  {Renzini}, {Stern}, \& {GOODS Team}}]{Vanzella2008}
{Vanzella} E. {et~al.}, 2008, A\&A, 478, 83

\bibitem[{{Whitaker} {et~al}\mbox{.}(2014){Whitaker}, {Franx}, {Leja}, {van
  Dokkum}, {Henry}, {Skelton}, {Fumagalli}, {Momcheva}, {Brammer}, {Labb{\'e}},
  {Nelson}, \& {Rigby}}]{Whitaker2014}
{Whitaker} K.~E. {et~al.}, 2014, ApJ, 795, 104

\bibitem[{{Whitaker} {et~al}\mbox{.}(2012){Whitaker}, {van Dokkum}, {Brammer},
  \& {Franx}}]{Whitaker2012}
{Whitaker} K.~E., {van Dokkum} P.~G., {Brammer} G., {Franx} M., 2012, ApJL,
  754, L29

\end{thebibliography}
\bibliographystyle{mn2e}

\end{document}